%% file: Paper.tex
\documentclass[journal]{IEEEtran}
\usepackage{nicematrix,inputenc}
\usepackage{graphicx}
\usepackage{color}
\usepackage{epsfig,epstopdf} 
\usepackage[export]{adjustbox}
\usepackage{amssymb,bbm}  
\usepackage{authblk}
\usepackage{cite}
\usepackage{hyperref}
\usepackage{cleveref,multicol}
\usepackage{caption}
\usepackage{subcaption,float,array,arydshln}
\usepackage{array}
\usepackage{epsf}
\usepackage{times}
\usepackage{bigstrut}
\usepackage{amsmath}
\usepackage{amsxtra}
\usepackage{amsthm}
\usepackage{algorithmic}
\usepackage{algorithm}
\usepackage{xurl}
\usepackage{comment}
\usepackage{siunitx}
\usepackage{lipsum}
\usepackage{kantlipsum}
\usepackage{dblfloatfix}
\usepackage{adjustbox}
\usepackage{tabularx}
\usepackage{float,xcolor}
\usepackage{framed}
\usepackage{caption}
\newtheorem{definition}{\textbf{\text{Definition}}}
\newtheorem{theorem}{\textbf{\text{Theorem}}}
\newtheorem{lemma}{\textbf{\text{Lemma}}}

\newtheorem{corollary}{Corollary}

\newtheorem{remark}{Remark}
\crefrangelabelformat{equation}{(#3#1#4--#5#2#6)}
\crefname{equation}{Eq.}{Eqs.}
\Crefname{equation}{Equation}{Equations}
\renewcommand\IEEEkeywordsname{Keywords}
\usepackage{tikz}
\usepackage{tikzscale,booktabs,cancel}
\usepackage{adjustbox}
\usepackage{blkarray, bigstrut} 
\usetikzlibrary{positioning,backgrounds,decorations.text}
\newcommand{\X}[0]{\ensuremath{\boldsymbol{X}}}
\newcommand{\A}[0]{\ensuremath{\boldsymbol{A}}}
\newcommand{\E}[0]{\ensuremath{\boldsymbol{E}}}
\newcommand{\D}[0]{\ensuremath{\boldsymbol{D}}}
\newcommand{\I}[0]{\ensuremath{\boldsymbol{I}}}
\newcommand{\Q}[0]{\ensuremath{\boldsymbol{Q}}}
\renewcommand{\H}[0]{\ensuremath{\boldsymbol{H}}}
\renewcommand{\P}[0]{\ensuremath{\boldsymbol{P}}}
\renewcommand{\o}[0]{\ensuremath{\boldsymbol{0}}}
\newcommand{\QG}[0]{\ensuremath{\boldsymbol{QG}}}
\newcommand{\QU}[0]{\ensuremath{\boldsymbol{QU}}}
\newcommand{\QD}[0]{\ensuremath{\boldsymbol{QD}}}

\newcommand{\HU}[0]{\ensuremath{\boldsymbol{HU}}}
\newcommand{\HD}[0]{\ensuremath{\boldsymbol{HD}}}

\newcommand{\V}[0]{\ensuremath{\boldsymbol{\rm V}}}

\newcommand\Tstrut{\rule{0pt}{2.4ex}}         % = `top' strut
\input{defs_tikzpgf}
\usetikzlibrary{petri}
\usepackage{tikzscale}

\newcommand{\Cross}{$\mathbin{\tikz [x=1.4ex,y=1.4ex,line width=.2ex, black] \draw (0,0) -- (1,1) (0,1) -- (1,0);}$}%
\begin{document}
% \title{Time and Energy Constrained Large-Scale IoT Networks: The Feedback Dilemma}

% \title{Open/Closed-loop Rate Adaptation for Time and Energy Constrained Large-Scale IoT Networks}
\title{Rate Adaptation in Delay-Sensitive and Energy-Constrained Large-Scale IoT Networks}

\author{Mostafa Emara,~\IEEEmembership{Member,~IEEE,}
        Nour~Kouzayha,~\IEEEmembership{Member,~IEEE,}
		Hesham~ElSawy,~\IEEEmembership{Senior Member,~IEEE,}
		and~Tareq Y. Al-Naffouri,~\IEEEmembership{Senior Member,~IEEE}
		\thanks{M. Emara, N. Kouzayha, and T. Y. Al-Naffouri are with King Abdullah University of Science and Technology, Thuwal, Saudi Arabia (e-mail: mostafa.emara@kaust.edu.sa; nour.kouzayha@kaust.edu.sa; tareq.alnaffouri@kaust.edu.sa).} 
		\thanks{H. ElSawy is with the School of Computing, Queen's University, Ontario, Canada (e-mail: hesham.elsawy@queensu.ca).}
  \vspace{-0.2cm}
		%\thanks{Manuscript received xxxx, xxxx; revised xxxx, xxxx.}
	}

\maketitle
%%%%%%%%%%%%%%%%%%%%%%%%%%%%%%%%%%%%%%%
\begin{abstract}
% Novel Internet of Things (IoT) access technologies have emerged as a part of next-generation networks. Transmission reliability, packet delivery latency, and energy consumption are key performance indicators (KPIs) in large-scale IoT networks. Rate adaptation based on packet fragmentation is an agile approach to improve transmission reliability. Rate adaptation error control depends on the availability of feedback, which is necessary to maintain efficient and reliable wireless links. Meanwhile, imperfect (error-prone) feedback negatively impacts the transmission reliability. In the 6G era, IoT devices may not be able to support feedback transmissions due to stringent energy constraints. This calls for new transmission techniques and design paradigms to maintain reliability in feedback-free IoT networks. In this context, this paper introduces transmission schemes based on whether feedback is present or not, namely, closed-loop rate adaptation (CLRA) and open-loop rate adaptation (OLRA).

Feedback transmissions are used to acknowledge correct packet reception, trigger erroneous packet re-transmissions, and adapt transmission parameters (e.g., rate and power). Despite the paramount role of feedback in establishing reliable communication links, the majority of the literature overlooks its impact by assuming genie-aided systems relying on flawless and instantaneous feedback. An idealistic feedback assumption is no longer valid for large-scale Internet of Things (IoT), which has energy-constrained devices, susceptible to interference, and serves delay-sensitive applications. Furthermore, feedback-free operation is necessitated for IoT receivers with stringent energy constraints. In this context, this paper explicitly accounts for the impact of feedback in energy-constrained and delay-sensitive large-scale IoT networks. We consider a time-slotted system with closed-loop and open-loop rate adaptation schemes, where packets are fragmented to operate at a reliable transmission rate satisfying packet delivery deadlines. In the closed-loop scheme, the delivery of each fragment is acknowledged through an error-prone feedback channel. The open-loop scheme has no feedback mechanism, and hence, a predetermined fragment repetition strategy is employed to improve transmission reliability. Using tools from stochastic geometry and queueing theory, we develop a novel spatiotemporal framework to optimize the number of fragments for both schemes and repetitions for the open-loop scheme. To this end, we quantify the impact of feedback on the network performance in terms of transmission reliability, latency, and energy consumption.

\end{abstract}


%%%%%%%%%%%%%%%%%%%%%%%%%%%%%%%%%%%%%%%%%%%%%%%%%%%%%%
\vspace{-0.3cm}
\section{Introduction} \label{Intro}
%The heterogeneity of the IoT use cases, along with the massive numbers of devices, call for novel transmission techniques to accommodate such unrelenting traffic demand while satisfying the diverse quality of service (e.g., reliability, latency, energy consumption, etc.) constraints~\cite{bennis2018ultrareliable,jurdi2018outage}. 
The fifth generation (5G) and beyond wireless systems are foreseen to support massive Internet of Things (IoT) deployments~\cite{zhang20196g,saad2019vision}. This is evident by an unprecedented proliferation of IoT devices, which are expected to exceed $5$~billion by $2025$ to enable ubiquitous monitoring and smart automation of industrial systems, precision agriculture, intelligent transportation, remote healthcare, and public safety verticals~\cite{Ericsson_Report}. Many of the emerging IoT use cases target delay-sensitive applications, where generated packets should be delivered within a predefined hard deadline~\cite{bennis2018ultrareliable, jurdi2018outage, tang2021spatiotemporal}. Otherwise, the information within the packet becomes obsolete and is not worth transmission. For a reliable transmission that fulfills such delay constraints, rate adaptation via packet fragmentation and repetition is widely adopted in IoT technologies such as Narrow-band IoT (NB-IoT)~\cite{liberg2017cellular} and Long-range wide area networks (LoRaWAN)~\cite{yegin2020lorawan,aguilar2021evaluation}.

In time-slotted systems, having smaller fragments implies reduced transmission rate, which in turn increases reliability of transmission per fragment\footnote{Instances of time-slotted multiple access are prevalent across various contemporary wireless technologies such as NB-IoT, LoRa, Sigfox, Zigbee, Bluetooth, among others. However, we maintain a general discussion to avoid diverting into intricate technology-specific nuances that might detract from the paper's focus and diminish its significance without enhancing the treatise.}. %Examples of time-slotted multiple access can be found in many of the current wireless technologies (e.g., NB-IoT, LoRa, Sigfox, Zigbee, Bluetooth, etc). Nevertheless, we keep our discussion general to avoid delving into cumbersome technology-specific details that will distract the focus of the paper and dilute its contributions without benefiting the treatise.} 
However, the trade-off is that smaller fragments require a larger number of successful transmissions to send a packet~\cite{elsawy2020rate, Nabil_adaptation}. In the presence of acknowledgment feedback, ensuring reliability and transitioning from one fragment to the next is straightforward since successfully transmitted fragments are acknowledged. However, in feedback-free scenarios, the transmitter lacks information about fragment delivery status. Consequently, it resorts to sending multiple copies of each fragment across various time slots to increase the chances of successful delivery~\cite{zhao2018deploying,masoudi2018grant}. In addition to acknowledging the transmission status of fragments, feedback transmission creates an online close-loop mechanism for rate adaptation and power control~\cite{huang2020wireless}, which is absent in the open-loop feedback-free counterpart~\cite{yu2017uplink,zheng2021open}.  Feedback-free operation is foreseen to dominate IoT applications with stringent energy constraints. Being a major source of energy consumption, wireless transmissions should be minimized where applicable to conserve the scarce energy resource in IoT networks. In this context, eliminating the feedback transmission might be a sought solution to conserve the energy of IoT receivers, which reduces the overwhelming burden to monitor, recharge, and/or replace their batteries~\cite{khodr2017energy,kouzayha2017joint}. Nevertheless, the fundamental role of feedback in wireless systems calls for innovative solutions to counter the impact of its absence, which is important to balance the trade-off between energy conservation and IoT network performance in terms of reliability and latency.

Motivated by the fundamental role of feedback on emerging delay-sensitive and energy-constrained IoT applications, this paper investigates closed-loop (i.e., feedback-assisted) and open-loop (i.e., feedback-free) rate adaptation schemes in large-scale networks. To design and assess the network key performance indicators (KPIs) of both schemes, we develop a novel spatiotemporal mathematical framework using tools from stochastic geometry and queuing theory. For the closed-loop scheme, rate adaptation is adopted based on feedback acknowledgments transmitted from the IoT receivers to their intended transmitters. In contrast to the common assumption of a perfect (lossless) delay-free feedback, we consider an error-prone feedback channel to investigate the negative impact of feedback impairments on transmission reliability and latency. On the other side, the open-loop scheme adopts transmitting multiple copies of each fragment aiming to increase the chance of successful delivery. Furthermore, the open-loop scheme is compared to the closed-loop benchmark. Our numerical results highlight the necessity of optimized packet fragmentation to attain a successful transmission within the predefined hard-deadline. By comparing the closed-loop and open-loop schemes, the impact of feedback presence/absence on the KPIs is revealed and quantified. Moreover, the trade-offs between transmission reliability, latency, and energy consumption are characterized.

%Utilizing both packet fragmentation and repetition for reliable transmissions in large-scale IoT networks, this paper introduces innovative transmission schemes that revolve around the availability or absence of feedback: CLRA and OLRA transmission schemes. 

\subsection{Related Work}
% {\color{blue} We need to talk here about IoT energy conservation (time?) works and how they were studied, then move to talk about spatial-temporal.}

Developing mathematical frameworks to analyze the performance of large-scale time-slotted IoT networks is an attractive research topic that has been widely studied in recent years ~\cite{kouzayha2017joint,elsawy2016modeling,bader2017first,benkhelifa2020recycling,abd2018joint,emara2020prioritized}. Spatiotemporal models are considered to jointly account for the massive spatial existence of IoT devices and the sporadic traffic flow per device~\cite{zhong2017heterogeneous,gharbieh2017spatiotemporal,zhang2021spatiotemporal,bader2017first}. From the temporal perspective, queuing theory is utilized to conduct the microscopic analysis that accounts for the packets departure/arrival along with the devices activities. From the spatial perspective, the macroscopic analysis relies on stochastic geometry that models the aggregate interference among active devices operating on the same channel due to the shared nature of the wireless channel~\cite{elsawy2016modeling}. 

The recently developed spatiotemporal models have gained popularity in characterizing the transmission reliability in terms of packet successful delivery probability, latency, scalability, and stability of large-scale IoT networks~\cite{gharbieh2017spatiotemporal, gharbieh2018spatiotemporal, jiang2018analyzing, chisci2019uncoordinated, moussa2018rach, yang2019spatio, yang2018delay}. For instance, scalability and stability of random access in IoT networks are characterized in~\cite{gharbieh2017spatiotemporal,jiang2018analyzing, chisci2019uncoordinated, moussa2018rach}. Transmission latency of the downlink IoT network is evaluated in~\cite{yang2019spatio, yang2018delay}. A spatiotemporal analysis of an uplink network is conducted in~\cite{gharbieh2018spatiotemporal}. However, the studies in~\cite{bader2017first,gharbieh2017spatiotemporal, gharbieh2018spatiotemporal, jiang2018analyzing, chisci2019uncoordinated, moussa2018rach, yang2019spatio, yang2018delay} ignore packet latency constraints (i.e., transmission deadlines), which are critical for delay-sensitive IoT applications. The authors in~\cite{elsawy2020characterizing,li2020age,tang2021spatiotemporal} account for hard-packet deadlines while considering asynchronous periodic traffic in~\cite{elsawy2020characterizing}, multi-cast traffic in~~\cite{li2020age} and multi-stream traffic in~\cite{tang2021spatiotemporal}, respectively. However, none of the spatiotemporal models in~\cite{bader2017first,gharbieh2017spatiotemporal, gharbieh2018spatiotemporal, jiang2018analyzing, chisci2019uncoordinated, moussa2018rach, yang2019spatio, yang2018delay, elsawy2020characterizing,li2020age,tang2021spatiotemporal} applies packet fragmentation nor time diversity (i.e., packet repetition) techniques to improve the transmission reliability.       

Among the developed spatiotemporal models that characterize IoT networks, specific interest is devoted to highlighting the gains and trade-offs of the rate adaptation and repetition techniques in terms of extended coverage and increased latency. For instance, the spatiotemporal model proposed in~\cite{elsawy2020rate} highlights the effect of static and dynamic rate adaptation in IoT networks. The works in~\cite{she2017cross,wu2019urllc} characterize the delay and reliability in ultra-reliable and low-latency (URLLC) IoT networks with time diversity. However, the packet deadline constraints are ignored in~\cite{elsawy2020rate,she2017cross,wu2019urllc}. {For delay-constrained networks, frame repetition is applied in~\cite{manzoor2021iot,manzoor2020optimal, manzoor2022improving} to improve the successful packet delivery. However, the analysis in \cite{manzoor2021iot,manzoor2020optimal, manzoor2022improving} does not explicitly capture the temporal aspects of the traffic and KPIs. }

The aforementioned spatiotemporal models assume that the feedback is lossless (perfect) and instantaneous (delay-free), which overlooks the fundamental impact of the inevitable feedback channel impairments. The work in ~\cite{rezasoltani2021real} studies the impact of the erroneous feedback channel on the age of information performance for a point-to-point (i.e., not large-scale network) scenario. In fact, none of the available studies in the literature have developed a spatiotemporal analytical framework for delay-sensitive and energy-constrained large-scale IoT networks with imperfect feedback channels and open-loop/closed-loop rate adaptation (i.e., via fragmentation and repetition), which is the main focus of this paper.
\vspace{-5pt}
\subsection{Contributions and Organization}
This article provides a novel spatiotemporal framework that utilizes stochastic geometry and queuing theory to characterize the transmission reliability, latency, and energy consumption trade-off of the open-loop rate adaptation (OLRA) and closed-loop rate adaptation (CLRA) schemes in delay-sensitive and energy-constrained large-scale IoT networks. An absorbing Markov chain (MC) is utilized to capture the temporal dimension and model the traffic flow of the IoT network with packet deadline constraint. Extensive Monte Carlo simulations are conducted to validate the accuracy of the developed analytical models. To the best of the authors' knowledge, this is the first work that provides a spatiotemporal mathematical analysis of large-scale IoT networks with rate adaptation, repetition, and feedback imperfections. The main contributions of this paper are summarized as follows. 
\begin{itemize}
\item We develop a spatiotemporal model to characterize a feedback-free OLRA scheme with packet fragmentation and repetition to improve the likelihood of successful packet delivery. The proposed OLRA scheme is optimized and benchmarked with the CLRA scheme with packet fragmentation and feedback transmission. 
%\item We develop a novel spatiotemporal model to characterize the OLRA scheme in which the packet is divided into equal-size fragments. Due to the absence of feedback, packet fragments are sent multiple times, irrespective of the decoding process status at the test receiver. Two different schemes, namely, OLRA with fixed rate (OLRA-FR) and OLRA with variable rate (OLRA-VR) are considered.
\item  We explicitly account for feedback channel impairment in the CLRA scheme in which the receiver acknowledges the status (success or failure) of the transmitted fragments. The feedback-error-free CLRA scheme is used to benchmark the feedback-error-prone CLRA as well as the proposed OLRA schemes. %Unlike conventional assumptions of a perfect delay-free feedback channel in existing literature, we consider an error-prone feedback channel to emphasize the impact of feedback impairments on packet transmission reliability and latency. 
%\item Using the matrix analytical method (MAM)~\cite{alfa2016applied}, we express mathematically the transmission reliability defined by the packet success delivery probability, the transmission latency, and the energy consumed by the receiver in packet decoding for both the CLRA and OLRA schemes.
 %We conduct extensive Monte-Carlo simulations and validate the accuracy of the derived expressions. 
\item We highlight and quantify the impact of the presence/absence of feedback and the imperfection of the feedback channel in terms of reliability, latency, and energy consumption. We also underscore the superiority of OLRA in reducing the energy consumption of IoT devices when compared to the CLRA scheme. 
\end{itemize}
%The analysis is based on the well-celebrated matrix analytical method (MAM)~\cite{alfa2016applied}, which is used to we express mathematically the transmission reliability defined by the packet success delivery probability, the transmission latency, and the energy consumed by the receiver in packet decoding for both the CLRA and OLRA schemes. Moreover, the effects of the presence/absence of feedback and the imperfection of the feedback channel are investigated.

The rest of this paper is organized as follows. Section~\ref{sec: system model} introduces the system model. The CLRA and OLRA transmission schemes are defined in Section~\ref{sec: MAC schemes}. Section~\ref{sec: analysis} presents the temporal, spatial, and performance metrics analysis for the considered transmission schemes. Section~\ref{sec: numerical results} explains the numerical results. Finally, the paper is concluded in Section~\ref{sec: conclusion}.  
\section{System Model}
\label{sec: system model}
\subsection {Spatial Parameters} \label{subsec: spatial parameters}
We focus on a pair of IoT transmitter/receiver (Tx/Rx) devices separated by a distance $R_o$. The impact of other coexisting IoT devices is captured via a heterogeneous Poisson field (HPF) of interferers, which is modeled by a marked Poisson process $(\Psi,\V)$. The locations of the interfering IoT devices are abstracted by a Poisson point process (PPP) $\Psi$ of intensity $\lambda$. A set of marks $\V= \{1,2,\cdots,V\}$ of an arbitrary density function $f_{\rm v}(v)$, independent of the devices' locations, are used to reflect the different types of the coexisting IoT devices. Each device of mark $v \in \V$ has a transmission power $p_v $ and an activity factor $\alpha_v$. Hence, the IoT device is active and can interfere with the intended transmission with probability $\alpha_v$ and is idle with probability $1-\alpha_v$. The locations and types of the interfering devices are assumed static once realized due to the short time slot assumption that prevents tangible changes in the locations or types of devices. 

Without loss of generality, a test receiver located at the origin is considered to analyze the network performance. The power-law path loss model is assumed in which the power of transmitted signals decays with the distance $r$ at the rate $r^{-\eta} $, where $\eta>2$ is the path-loss exponent. Moreover, the Rayleigh fading channel of unit mean power fading is considered, which is independent of different locations and types of IoT devices.

%%%%%%%%%%%%%%%%%%%%%%%%%%%%%%%%%%%%%%%%%%%%%%%%%%%%%%%%%%%%
\begin{figure*}[htp!]
\centering
% \begin{adjustbox}{width=0.9\textwidth,height=14cm}
% \input{SM2.tex}
% \end{adjustbox}
\includegraphics[width=\linewidth,height=14cm]{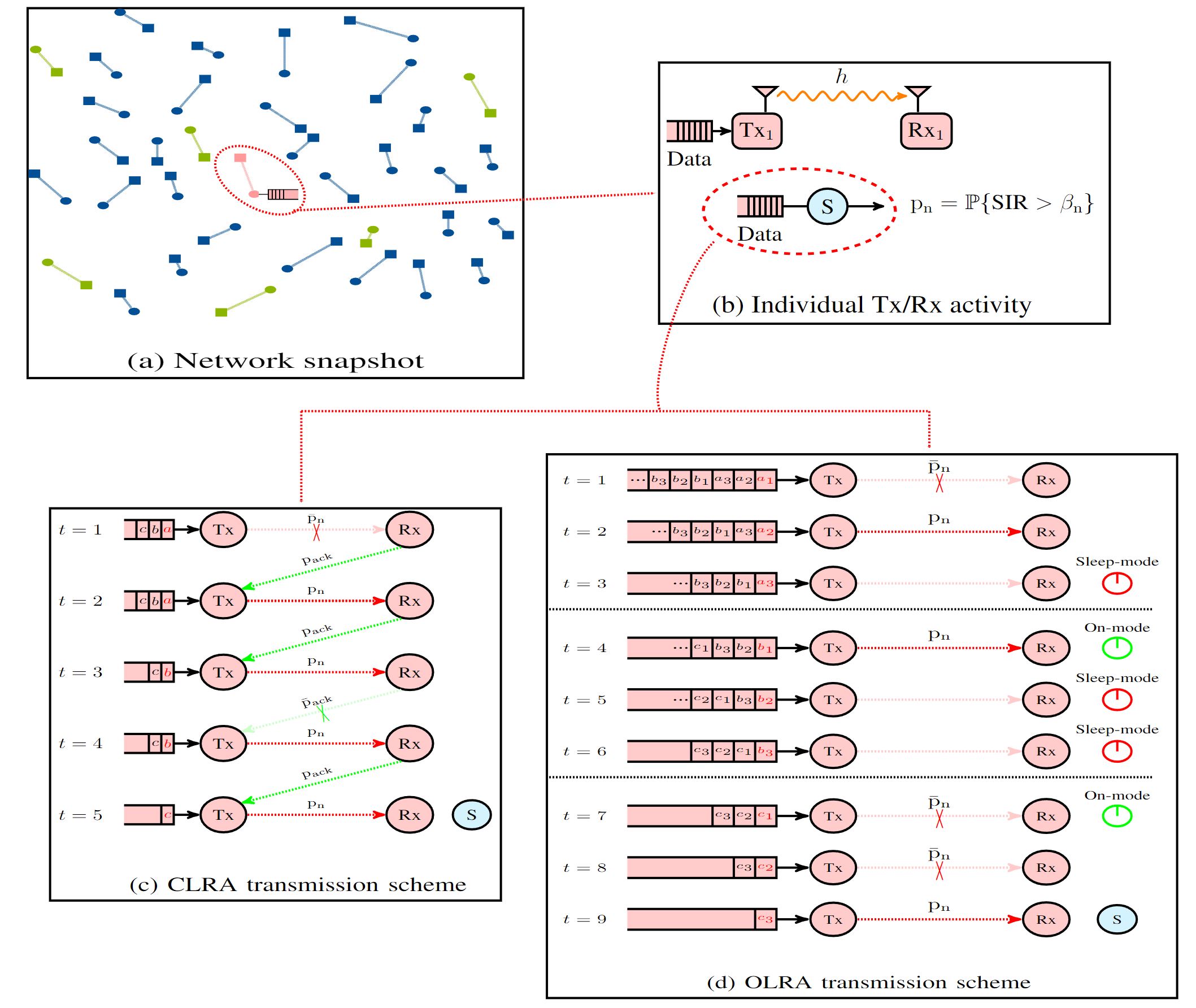}
\small \caption{(a) Snapshot of the network. Nodes, squares, and dashed lines represent transmitters, receivers, and Tx/Rx links, respectively. The test Tx/Rx is red-colored and surrounded by HPF of interferers where the active Tx/Rx links are blue-colored and inactive ones are green-colored. (b) The activity of the test Tx/Rx link. The test transmitter has a buffer for the packet fragments. The fragment is successfully delivered with probability $\rm p_n$. (c) CLRA transmission scheme in the case of packet successful delivery. (d) OLRA transmission scheme in the case of packet successful
delivery. Considering a packet of $n = 3$ fragments, indexed by $\{a,b,c\}$, and fragment repetition of $3$ times, indexed by $\{x_1, x_2, x_3\}$ for $x \in \{a,b,c\}$.}
\label{fig: system model}
\vspace{-0.3cm}
\end{figure*}
%%%%%%%%%%%%%%%%%%%%%%%%%%%%%%%%%
\subsection{Temporal Parameters and Main Performance Metrics} 
\label{subsec: temporal parameters} 
A time-slotted system is considered with deterministic traffic arrival, where a packet is generated every $T$ time slots. Transmitters with non-empty buffers are required to send packets of length $L$ bits with transmission rate $R_n$. Hence, packets are divided into $n \leq T$ equal fragments of length $\lceil \frac{ L}{n T_s} \rceil$, where the transmission and decoding of a fragment occur within a single time slot of duration $T_s$. The transmission rate $R_n$ can be expressed as
\begin{equation}
	R_n=\frac{L}{n T_s}=W \log_2 (1+\theta_n), 
	\label{rate}
\end{equation}
where $W$ is the frequency bandwidth and $\theta_n=2^{\frac{R_n}{W}}-1=2^{\frac{L}{n W T_s}}-1$ is the signal to interference ratio (SIR) detection threshold required to correctly decode the fragment at the intended receiver. A packet due-time of $T$ time slots is assumed to represent hard transmission deadlines\footnote{Such assumption is convenient for time-constrained IoT applications that require \emph{fresh} updates or measurements.}. Hence, at $t=T$, the packet is dropped from the transmitter buffer whether it is successfully delivered or not. The packet success delivery (PSD) is achieved by correctly decoding all fragments within the packet deadline $T$. A fragment is successfully delivered by the test receiver if the received SIR is larger than the detection threshold $\theta_n$. Thus, the fragment success delivery (FSD) probability, denoted by $\rm p_n$, is defined as $\rm p_n = \mathbb{P} \left \{\text{SIR} \geq \theta_n \right\}$. 

According to (\ref{rate}), dividing the packet into more fragments enhances the transmission reliability as it leads to a lower detection threshold $\theta_n$ that is more likely to be satisfied. This, however, expands the packet transmission over multiple time slots and may increase the overall packet delivery latency. In this work, we adopt the \emph{\textbf{PSD probability}} and the \emph{\textbf{PSD mean latency}} as the main performance metrics to assess and compare the different proposed schemes. Moreover, the \emph{\textbf{energy consumption}} of the IoT receiver during the decoding process is also considered to highlight the impact of feedback and fragment repetition on energy-constrained IoT networks.

\section{Transmission Schemes}\label{sec: MAC schemes}
To enhance the PSD probability of the IoT network, we consider a fragmentation and repetition mechanism, according to which each fragment is transmitted several times. The repetition mechanism is defined based on the adopted transmission policy and whether the feedback exists or not. The CLRA is a reactive scheme, where retransmissions are triggered by a feedback signal that acknowledges the status of the fragment decoding at the receiver. In the OLRA scheme, the feedback is absent, and hence, the transmitter proactively decides on the number of repetitions to maximize the PSD probability. The transmission/decoding procedure for each scheme is detailed in the sequel.
\vspace{8pt}
\subsubsection{\textbf{Closed Loop Rate Adaptation (CLRA) Scheme}}  
\label{subsubsec: CLRA definition}
In this scheme, the intended receiver attempts to decode the transmitted fragment. In the case of successful decoding, the receiver sends an acknowledgment (ACK) message via a control channel to the transmitter to drop this fragment from its queue and send the subsequent fragment in the next time slot. Otherwise, the receiver sends a negative ACK (NACK) asking for the retransmission of the fragment. The transmitter in the NACK case keeps the same fragment at the head of its queue and persists in sending it until receiving a positive ACK from the receiver or reaching the maximum number allowed for retransmissions. An error-prone (imperfect) control channel is considered to emphasize the impact of feedback impairments on the performance metrics. Let $\rm p_{ack}$ denote the probability that the feedback acknowledgment (ACK/NACK) sent by the test receiver is correctly received at the paired transmitter. Similarly to the FSD, $\rm p_{ack}$ is the probability that the received feedback SIR is larger than an acknowledgment detection threshold $\theta_{\text{ack}}$ (i.e., $\rm p_{ack} = \mathbb{P} \left \{{SIR_{ack}} \geq \theta_{\text{ack}} \right\} $). The detection threshold $\theta_{\text{ack}}=2^{\frac{L_{\text{ack}}}{W T_{\text{ack}}}}-1$, where $L_{\text{ack}}$ and $T_{\text{ack}}$ denote the acknowledgment message length and duration, respectively. Therefore, the successful delivery of a generic packet sent by $R_n$ transmission rate from a generic IoT device is jointly controlled by the fragments and feedback successful delivery probabilities $\rm p_n$ and $\rm p_{ack}$.
% Therefore, the successful delivery of a fragment is conditioned by jointly decoding success at the test receiver (of probability $\rm p_{n,m}$) and acknowledgement success receiving at the paired transmitter (of probability $\rm p_{ack}$). Hence, the fragment successful delivery probability of CLRA scheme is $\rho=\rm p_{n,m} \rm p_{ack}$.} 
The PSD and failure (time elapsed) events of the CLRA scheme are defined as follows. 			
\begin{definition}[CLRA PSD] 
A packet sent at rate $R_n$ is successfully delivered if the receiver correctly decodes and acknowledges all the $n$ fragments within the $T$ time-slots period.
\label{def:CLRA success}
\end{definition}
\begin{definition}[CLRA Packet Delivery Failure] At any instant~$t$, if the remaining time slots are insufficient to complete the transmission of the pending fragments related to the same packet, a CLRA failure event occurs. To save energy, a sleeping trigger signal is sent by the receiver to the transmitter to drop the packet, and both of them switch to sleep mode until the next packet generation.
\label{def:CLRA failure}
\end{definition}

\subsubsection{\textbf{Open Loop Rate Adaptation (OLRA) Scheme}} \label{subsubsec: OLRA definition} 
In this scheme, the transmitter lacks knowledge about the decoding status at the receiver, and hence, a predefined number of copies of the same fragment are transmitted to exploit temporal diversity. For the sake of maximizing the PSD, we assume that the transmitter exploits all the available $T$ time slots for fragment repetition. %\footnote{ This comes at the cost of higher energy consumption at the transmitter while relieving the receiver from sending feedback messages.} 
 This also simplifies the OLRA design as we only need to determine the optimal number of fragments~$n$, which subsequently determines the repetition count~$\kappa$. Specifically, each fragment is sent $\kappa=\lfloor T/n \rfloor $ times, where $n$ is the total number of fragments. For the remaining $\tau= \mod(T,n)$ time slots, the transmitter randomly selects $\tau <n$ fragments to be sent one more time each. Therefore, each of the selected fragments is sent $\kappa+1$ times while the others are repeated $\kappa$ times. Note that all fragments are sent with transmission rate $R_n$. The PSD and failure events of the OLRA schemes can be defined as follows.
\begin{definition}[OLRA PSD] A packet sent at a transmission rate $R_n$ is successfully delivered if at least one copy of each fragment is correctly decoded within the packet deadline $T$.
\label{def:OLRA success}
\end{definition}
\begin{definition}[OLRA Packet Delivery Failure] In the case of decoding failure for all copies of any fragment, an OLRA failure event occurs. To save energy, the receiver stops decoding the subsequent fragments and switches to sleep mode to conserve energy. It is worth mentioning that the transmitter does not switch to sleep mode since it is unaware of the receiver status.
\label{def:OLRA failure}
\end{definition}

Fig.~\ref{fig: system model} depicts a network snapshot in which nodes represent the IoT transmitters, squares denote the receivers, and dashed lines are for IoT Tx/Rx links. The test Tx/Rx link is represented in red color with a buffer at the test transmitter to store the packet fragments. The HPF set of interferers are blue-colored and green-colored for active and inactive Tx/Rx links, respectively. A fragment is successfully decoded with probability $\rm p_n$ that depends on the interference experienced at the test receiver. The first input first out (FIFO) discipline is assumed for packet service. Fig.~\ref{fig: system model} also offers a pictorial illustration of the transmission policies for CLRA and OLRA schemes in the case of packet successful delivery. The figure shows a packet of $n=3$ fragments, indexed by $\{\text{a,b,c}\}$ and fragment repetition of $3$ times, indexed by $\{x_1,x_2,x_3\}$ for $x \in \{\text{a,b,c}\}$.

The CLRA and OLRA schemes can be modeled using discrete-time absorbing Markov chains (MC) with two absorbing states, namely \emph{success} and \emph{timeout (failure)} states. The absorbing MC is fully characterized by the transition matrix $\P$ that tracks the decoding attempts of the fragments of a packet until it is eventually absorbed to either the success or failure states. The transition matrix $\P$ depends on the FSD probability $\rm p_n$ for the OLRA scheme and on $\rm p_n$ and the acknowledgment success probability $\rm p_{ack}$ for the CLRA scheme. We consider the different time scales of variation for the network traffic (i.e., packet generation and transmission) and channel fading when compared to the network spatial topology. In particular, the HPF of interfering devices is assumed fixed once realized, however, the channel fading and traffic vary at the scale of time slot. To account for the different IoT network realizations, we consider the meta distribution of the FSD probability. We then group the realizations that would lead to an FSD probability of a range $\pm \frac{1}{2M}$ into the same class, denoted hereafter as FSD class, where $M$ is determined based on the needed accuracy. Thus, $\rm p_{n,m}$ denotes the FSD probability of a fragment sent from a device that belongs to the $m$-th FSD class with a transmission rate $R_n$. On the other side, we consider averaging over all Rx/Tx links (mean-field) to find the feedback signaling success probability $\rm p_{ack}$. Given $\rm p_{n,m}$ and $\rm p_{ack}$, the transition matrix $\P^{(m)}$ for each FSD class can be formulated for the CLRA and OLRA schemes.

\section{Analysis}
\label{sec: analysis}
This section develops the mathematical frameworks to evaluate the performance of the CLRA and OLRA schemes. We start with the temporal analysis of the absorbing MC that characterizes each scheme. Hence, we construct the transition matrix $\P_{\text{CLRA}}^{(m)}$ for a device belonging to the $m$-th FSD class in terms of the successful decoding and acknowledgment probabilities $\rm p_{n,m}$ and $\rm p_{ack}$ for the CLRA transmission scheme. We follow a similar approach for the OLRA scheme, where we construct the transition matrices $\P_{\text{OLRA}}^{(m)}$ in terms of the successful decoding probability $\rm p_{n,m}$. Using stochastic geometry, the macroscopic network spatial analysis is handled to characterize the SIR meta distribution of the FSD for the different $M$ classes and provide an expression for the FSD probability $\rm p_{n,m}$, $m\in\{1,2, ... M\}$ in addition to characterizing the reverse-link feedback SIR and the acknowledgment success probability $\rm p_{ack}$ expressions. Finally, with the aid of the matrix analytical method (MAM)~\cite{alfa2016applied}, the \emph{PSD probability}, \emph{packet delivery average latency}, and the \emph{receiver energy consumption} are mathematically obtained.
%%%%%%%%%%%%%%%%%%%%%%%%%%%%%%%%%%%%%%%%%%%%%
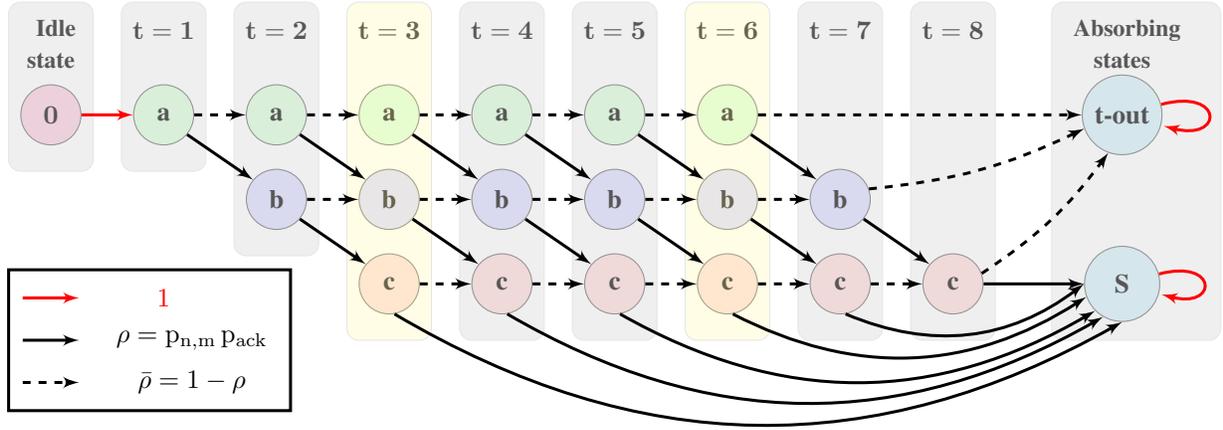
\begin{figure*}[h!]
	\centering
  \resizebox{0.9\linewidth}{!}{\input{MC_CLRA.tikz}}
    % \begin{adjustbox}{width=0.8\linewidth, height=5cm}
    % \input{MC_CLRA.tikz}
    %  \end{adjustbox}%
\small \caption{The absorbing MC of the CLRA scheme. The packet consists of $n=3$ fragments, denoted by $\{\text{a,b,c}\}$, and the packet deadline $T=8$.}
	\label{MC for CLRA}
 %\vspace{-0.5cm}
\end{figure*}
%%%%%%%%%%%%%%%%%%%%%%%%%%%%%%%%%%%%%%%%%%%%%
\subsection{Temporal Analysis}\label{subsec: temporal analysis} 
As previously mentioned, the considered schemes can be modeled by a discrete-time absorbing MC to track the fragments decoding attempts until reaching the final state where the packet is either successfully delivered or discarded due to the elapsed deadline. Consider a device belonging to the $m$th FSD class that transmits a packet with the rate $R_n$; its corresponding MC can be mathematically represented with the transition matrix $\P^{(m)}$ formulated as \cite[Section 3.6]{alfa2016applied}
\begin{equation}
\begin{aligned} 
	 \P^{(m)} & =	
	 \left[\arraycolsep=3.5pt\def\arraystretch{0.9}
		\begin{array}{c!{\vline width 1pt}c}
			\Q^{(m)} & \H^{(m)}\\
			\hline
			\o & \I\Tstrut
		\end{array} \right] 
	= 	\left[\arraycolsep=10pt\def\arraystretch{0.8}
	\begin{array}{c!{\vline width 1pt}c}
		\multicolumn{2}{c}{\tilde{\P}^{(m)}} \\ \hline
		\o & \I\Tstrut\\
	\end{array}
	\right] \\
		& = 
		\left[\arraycolsep=3.5pt\def\arraystretch{0.9}
		\begin{array}{c c c c  c !{\vline width 1pt} c}
			 \Q^{(m)}_1 & {\o} &  {\o} & \dots &  {\o} & \H^{(m)}_1\\[3pt]
			 {\o} & \Q^{(m)}_2 & {\o} & \dots &  {\o} & \H^{(m)}_2\\
		 \vdots & \ddots& \ddots & \ddots & \vdots& \vdots\\
		 {\o} & {\o} &  {\o} & \dots &  \Q^{(m)}_{T-2} & \H^{(m)}_{T-2}\\[3pt]
	  {\o} & {\o} &  {\o} & \dots &  \o &  \H^{(m)}_{T-1}\\[3pt]
			\hline
			  {\o} & {\o} &  {\o} & \dots &  \o &  \I		
		\end{array} \right],
		\unboldmath
\end{aligned}
\label{transition matrix}
\end{equation}
where $\Q^{(m)}$ denotes the transient matrix describing the attempts handled by the test receiver to decode the transmitted fragments before absorption, and $\H^{(m)}$ denotes the absorbing matrix and captures the probability that the packet is either successfully delivered or discarded due to elapsed deadline. $\I$ is an identity matrix of size $2\times 2$ representing the success and failure absorbing states. The rows in (\ref{transition matrix}) depict the progressing time evolution of the packet until absorption. The matrix $\Q^{(m)}_t$ represents the transition between time-slots $t$ and $t + 1$. $\H^{(m)}_t$ captures the absorption probability at time slot $t$. Next, we focus on formulating the matrix $\tilde{\P}^{(m)}$ presented in (\ref{transition matrix}) and consisting of only $\Q^{(m)} $ and $\H^{(m)}$ for the CLRA and OLRA schemes. Without notation abuse, we drop the subscripts in $\rm p_{n,m}$ and $\rm{\bar{p}_{n,m}}=1-\rm p_{n,m}$, in the hereafter matrices. 

\vspace{5pt}
\subsubsection{\textbf{Temporal Analysis of the CLRA Scheme}} \label{subsubsec: CLRA analysis}
To facilitate the exposition of the transition matrix $\tilde{\P}_{\text{CLRA}}^{(m)}$ of the CLRA scheme, we first present an illustrative example for the transmission of a packet with $n=3$ fragments and packet due-time $T=8$~time slots. We then generalize $\tilde{\P}_{\text{CLRA}}^{(m)}$ for different transmission rates and due-times. Fig.~\ref{MC for CLRA} depicts the absorbing MC of the considered illustrative example. The three fragments of the packet are denoted as $\{a,b,c\}$ and are retransmitted several times according to the feedback of the receiver. Fig.~\ref{MC for CLRA} starts with an idle state to represent the empty buffer at the test transmitter at instant $t=0$. At $t=1$, fragment $a$ is sent, and its first decoding attempt is handled at the test receiver. This is represented by a transition with probability $1$ between time slots $t=0$ and $t=1$. %It is worth noting that if an Aloha protocol with parameter $p_A$ is applied, the transmitter choices either to transmit with probability $p_A$ or to defer transmission to a subsequent time slot with probability $\bar{p}_A$. 
The successful delivery of a packet's fragment sent from an $m$-FSD class IoT device with a transmission rate $R_n$ is conditioned on the joint successful decoding at the test receiver (of probability $\rm p_{n,m}$) and successful acknowledgment (of probability $\rm p_{ack}$). Thus, the fragment success delivery probability of the CLRA scheme is $\rho=\rm p_{n,m} \rm p_{ack}$. If fragment $a$ is successfully delivered (i.e., jointly decoded and acknowledged) of probability $\rho$, the transmitter drops it from the queue and switches to transmit fragment $b$ in the next time slot. Otherwise, the transmitter persists in sending fragment $a$ during the subsequent time slots. The retransmission/decoding trials of fragment $a$ proceed until it is successfully delivered or the maximum number of retransmissions is reached. If the later event happens, the packet is discarded. Note that a packet is discarded at instant $t$ if the remaining $(T-t)$ time slots are insufficient to transmit all the pending fragments related to the same packet. When this happens, both transmitter and receiver switch to sleeping mode until the next packet generation. The same strategy is followed for the other fragments until the CLRA PSD event, defined in Definition~\ref{def:CLRA success}, occurs. According to this policy, the instant $t=2$ carries either the $2^\text{nd}$ trial to deliver fragment $a$ after a single failure (due to either decoding failure or corrupted/lost ACK) or the $1^\text{st}$ attempt to deliver fragment $b$ after the successful delivery of fragment $a$. Then, $t=3$ can have one of the following $4$ possibilities; (i) the $3$rd trial to deliver fragment $a$ after two consecutive failures; (ii) the $1$st trial to deliver fragment $b$ after one failure followed by a single success in delivering fragment $a$; (iii) the $2$nd delivering attempt of fragment $b$ after the successful delivery of fragment $a$ followed by a failure of delivering fragment $b$; and (iv) the $1$st trial to deliver fragment $c$ after two consecutive successful delivering attempts for fragments $a$ and $b$.

By construction, at least $n$ time slots are required for the CLRA packet delivery event to succeed. This is visualized in Fig.~\ref{MC for CLRA} by following the best-case scenario represented by the diagonal transitions from $a$ to $b$ then $c$, where the success event occurs, at $t=3$. On the other side, the packet discarding event is due to successive failures in delivering the transmitted fragments. This is also manifested in Fig.~\ref{MC for CLRA} by the absorption into the timeout state, for example, at $t=6$ after $6$ successive delivering failures of fragment $a$. The detailed structure of the transition matrix $\tilde{\P}_{\text{CLRA}}^{(m)}$ for the absorbing MC in Fig.~\ref{MC for CLRA} is given by (\ref{P-CLRA}). 
\begin{figure}
{\footnotesize
	\renewcommand\arraystretch{0.75}
 \makeatletter\setlength\BA@colsep{1.75pt}\makeatother
\begin{align}
&\tilde{\P}_{\text{CLRA}}^{(m)}  = \left[\begin{array}{c !{\vline width 1pt} @{\hspace{1ex}} c}
			\Q_{\text{CLRA}}^{(m)} & \H_{\text{CLRA}}^{(m)}
		\end{array} \right]\nonumber \\[3pt]
  %%%%%%%%%
		& = \quad \def\ra{\color{red}}
	\def\rb{\color{blue}}
	\begin{blockarray}{c @{\hspace{1cm}} c @{\hspace{1ex}} *2{c}| *3{c}| *3{c}| *3{c}| *3{c}| *2{c}| {c} |{c}!{\vline width 1.5pt}*2{c}}
 %%%
	\begin{block}{c @{\hspace{1cm}} c @{\hspace{1ex}} *2{c} *3{c} *3{c} *3{c} *3{c} *2{c} {c} {c} *2{c}}
		% \ra t & & 1 & {\hspace{1.5ex}}\ra 2 &  &	 & \ra 3 & & & \ra 4 &	& & \ra 5 & & & \ra 6 &	 & {\hspace{1.5ex}} \ra 7 &	 & \ra 8 & & \\
	 & &  \rb {a} & \rb {b} &	\rb {a} & \rb {b} &	\rb {c} &\rb {a} & \rb {b} &	\rb {c} &\rb {a} & \rb {b} &	\rb {c} &\rb {a} & \rb {b} &	\rb {c} & \rb {b} &	\rb {c} & \rb {c} & & \rb \text{S} & \rb \text{t-out} \\[-3pt]
\end{block}
 %%%%%%%%%%%%%%%%%%%%%%
\begin{block}{c @{\hspace{1ex}} c @{\hspace{1.5ex}} [*2{c}| *3{c}| *3{c}| *3{c}| *3{c}| *2{c}| {c}| {c}!{\vline width 1.5pt} *2{c}]}
		\ra t=1 & \rb {a} & \bar{\rho} & \rho &  &  &  &  &  &  &  &  &  &  &  &  &  &  &  &  &  &\bigstrut[b] \rule{0pt}{4.5ex} \\   \cline{1-22}
%%%%%%%%%%%%%%%%%%%%
	& \rb {a} &  & & \bar{\rho} & \rho & 0 &  &  &  & &  & &  &  & & & & & &  &  \Tstrut \\
		\ra t=2 & \rb {b}  & & & 0 & \bar{\rho}\ & \rho   & & & & & & & & & & & & & &  & \\	\cline{1-22}
% %%%%%%%%%%%%%%%%%%%%
& \rb {a} &  & &   &  &  & \bar{\rho} & \rho & 0 &  &  &  &  & &  &  & & & &  0 &  0 \Tstrut\\
\ra t=3 & \rb {b} &   & &  &  &   & 0 & \bar{\rho} & \rho &  &  &  &  &  & &  & & &  & 0 &  0 \\
& \rb {c} &  &  &  &  &   & 0 & 0 & \bar{\rho}  &  &  &  &  &  &  &  & && & {\rho} &  0\\\cline{1-22}
% %%%%%%%%%%%%%%%%%%%%
& \rb {a}  & & &   &  &  &  &  &  & \bar{\rho} & \rho & 0&  &  &   &    & &  & & 0 & 0 \Tstrut\\
\ra t=4 & \rb {b}    & &  & &  &   &  &  & & 0 &  \bar{\rho} & \rho   &  &   &  &  &   &   & &  0  & 0\\	
& \rb {c}  & &  & &  &   &  &  &   &  0 & 0  & \bar{\rho} &  & &  & &&   &  & {\rho} & 0 \\\cline{1-22}
% %%%%%%%%%%%%%%%%%%%%%%%%%%
& \rb {a}  & & &   &  &  &  &  &  & &  &  &  \bar{\rho} & \rho & 0 &    & &  & & 0 & 0 \Tstrut\\
\ra t=5 & \rb {b}  &  &  & &  &   &  &  & &  &  &   & 0 &  \bar{\rho} & \rho &  &   & & & 0 &  0 \\	
& \rb {c}  & &  & &  &   &  &  &   &  &  & & 0 & 0  & \bar{\rho} & & &  & &  {\rho} & 0 \\\cline{1-22}
% %%%%%%%%%%%%%%%%%%%%%%%%%%%%
& \rb {a} &  & &   &  &  &  &  &  &  & & &  &  &   &  \rho  & 0 & &  & 0 &  \bar{\rho} \Tstrut\\
\ra t=6 & \rb {b}  &  &  & &  &   &  &  & & &    &  &  &    &  & \bar{\rho} &  \rho &  & & 0 & 0 \\	
& \rb {c} & & &   &  &   &  &  &   &   &   &    &  & & & 0 &\bar{\rho} &  & & {\rho} & 0 \\\cline{1-22}
% %%%%%%%%%%%%%%%%%%%%%%%%%%%%
\ra t=7 & \rb {b} & &  & &  &   &  &  &   &   &   &    &  & & &  & &  \rho &  &  0 &  \bar{\rho}\Tstrut\\
& \rb {c}  &  &  & &  &   &  &  & &  &     &  &  &    &  &  &   & \bar{\rho} & & {\rho} & 0 \\\cline{1-22}
% %%%%%%%%%%%%%%%%%%%%%%%%%%
\ra t=8 & \rb {c} & &   &  &  &   &  &  &   &   &   &   &  &   &  &  &  &   & 0 & {\rho} & \bar{\rho}\Tstrut \bigstrut\\	
\end{block}
\end{blockarray}
\label{P-CLRA}
\end{align}	}
\vspace{-1.0cm}
\end{figure}
%%%%%%%%%%%%%%%%%%%%%%%%%%%%%%%%%%%%%%%%%%%%%%%%%%
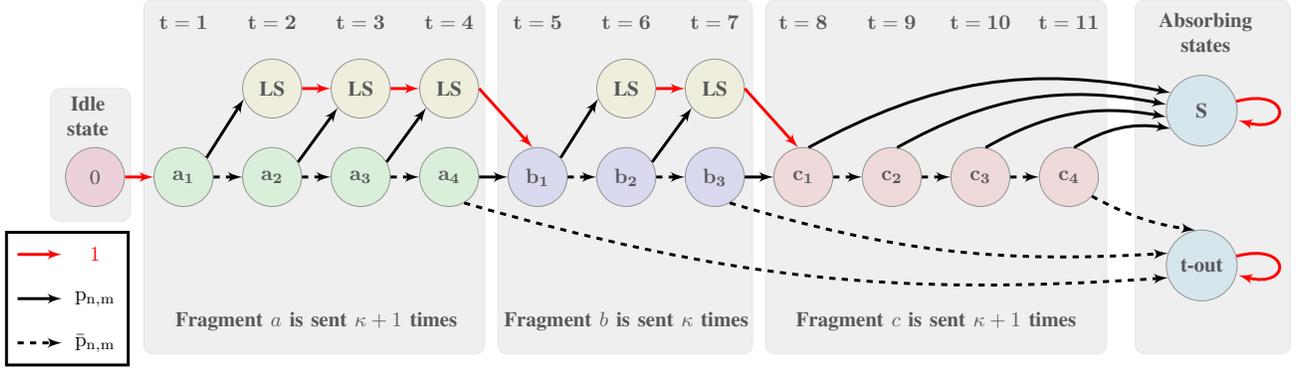
\begin{figure*}[h!]
	\centering
		\resizebox{0.95\textwidth}{!}{\input{MC_OLRA.tikz}}
	\small \caption{The absorbing MC of the OLRA scheme. The packet consists of $n=3$ fragments, denoted as $\{a,b,c\}$ and the packet deadline $T=11$. So, $\kappa=\lfloor T/n\rfloor = 3 $, and $\tau=\mod(T,n)=2$. Hence, each fragment is sent $\kappa$ times while $\tau=2$ fragments are randomly chosen to be sent one more time. We assume that $\{a,c\}$ are the selected fragments. The subscript in $x_i, x \in \{a,b,c\}$ denotes the $i$th decoding attempt.}
	\label{MC for OLRA}
 %\vspace{-0.5cm}
\end{figure*}
%%%%%%%%%%%%%%%%%%%%%%%%%%%%%%%%%%%%%%%%%%%%%%%%%%%%%%%%%%%%
The structure of $\tilde{\P}_{\text{CLRA}}^{(m)}$ in (\ref{P-CLRA}) follows the general transition matrix structure in (\ref{transition matrix}). The transient matrix $\Q_{\text{CLRA}}^{(m)}$ consists of the non-zero diagonal submatrices $\Q^{(m)}_t, t=\{1,2,\cdots, T-2\} $ and zero submatrices of appropriate sizes, which are left blanked. For $1 \leq t <n$, it can be noticed that $\Q^{(m)}_t$ is a fat matrix of dimension $t \times (t+1)$. Thus, its size gradually grows with time evolution. For $n \leq t < T-n+1$, $\Q^{(m)}_t$ is a square matrix with unchanged dimension of $n \times n$. Finally, for $T-n+1 \leq t \leq T-2$, $\Q^{(m)}_t$ is a thin matrix of dimension $(T-t+1) \times (T-t)$, hence it gradually declines with time progress. We denote by $\QG^{(m)}_t$, $\QU^{(m)}_t$ and $\QD^{(m)}_t$ the fat, square and thin $\Q^{(m)}_t$ matrices in the three mentioned time intervals. On the other hand, the absorbing matrix $\H_{\text{CLRA}}^{(m)}$ consists of two-column submatrices $\H_t^{(m)}, t=\{1,2,\cdots,T-1\}$ to capture the packet absorption into the success and failure absorbing states. It can be shown that the first possibility for a packet to be absorbed into the success state occurs at $t=n$ and into the timeout state at $t=T-n+1$, which complies with Fig.~\ref{MC for CLRA}. The transition matrix $\tilde{\P}_{\text{CLRA}}^{(m)}$ in (\ref{P-CLRA}) describing the absorbing MC of the CLRA scheme for $n=3$ fragments and packet due-time $T=8$. Lemma~\ref{Lemma: CLRA} provides the general CLRA transition matrix $\tilde{\P}_{\text{CLRA}}^{(m)}$ for arbitrary tuple $(n, T)$. 
\begin{lemma}[\textbf{CLRA Transition Matrix}] \label{Lemma: CLRA}
\normalfont The transition matrix $\tilde{\P}_{\text{CLRA}}^{(m)}$ describing the absorbing MC for a generic packet sent from an $m$-FSD class IoT device consists of the submatrices $\Q^{(m)}_{t,\text{CLRA}}$ and $\H^{(m)}_{t,\text{CLRA}}$ for an arbitrary tuple $(n, T)$, which are given by~\eqref{Q_CLRA} and~\eqref{H_CLRA}, respectively. $\X \Big|_{i \times j}$ denotes a matrix $\X$ of size $(i \times j)$ and the notations $\{G,U,D\}$ in $\{\QG_{t}^{(m)},\QU_{t}^{(m)},\QD_{t}^{(m)}\}$ and $\{\HU^{(m)}_{t}$, $\HD^{(m)}_{t}\}$ refer to \{growing, unchanged, and declining\}-size matrices, respectively. 
\begin{equation} \small
\Q^{(m)}_{t,\text{CLRA}}  = 
    \begin{cases}
		\QG_{t}^{(m)} \Big|_{t \times (t+1)}& 1 \leq t < n,\\[10pt]
		\QU_{t}^{(m)}\Big|_{n \times n} & n \leq t < T-n+1,\\[10pt]
		\QD_{t}^{(m)}\Big|_{(T-t+1) \times (T-t)} & T-n+1 \leq t \leq T-2,\\
	\end{cases}
 \label{Q_CLRA}
\end{equation}
\begin{equation}\small
\H^{(m)}_{t,\text{CLRA}} = 
    \begin{cases}
		\o \Big|_{t \times 2}& 1 \leq t < n,\\[10pt]
		\HU_{t}^{(m)}\Big|_{n \times 2} & n \leq t < T-n+1,\\[10pt]
		\HD_{t}^{(m)}\Big|_{(T-t+1) \times 2} & T-n+1 \leq t \leq T-2,\\[10pt]
		\begin{bmatrix}
			\rho & \bar{\rho}
		\end{bmatrix} & t=T-1,
	\end{cases} 
  \label{H_CLRA}
\end{equation}
 where the elements of the matrices $\QG_{t}^{(m)}$, $\QU_{t}^{(m)}$ and $\QD_{t}^{(m)}$  in~\eqref{Q_CLRA} are given by
 \begin{align}
 QG^{(m)}_{t} (i,j) &= QU^{(m)}_{t} (i,j) =\begin{cases}
 		\bar{\rho}, & \text{for } j=i,\\
 		\rho, & \text{for } j=i+1,\\
 		0, & \text{otherwise}, \\
 	\end{cases} \nonumber \\
QD^{(m)}_{t} (i,j) &=\begin{cases}
\rho, & \text{for } i=j,\\
 \bar{\rho}, & \text{for } i=j+1,\\
 0, & \text{otherwise}.
\end{cases}
\end{align}
and the elements of $\HU^{(m)}_{t}$ and $\HD^{(m)}_{t}$ in~\eqref{H_CLRA} are given by 
\begin{align}
	HU^{(m)}_t(i,j) &=
	\begin{cases}
		\rho, & \text{for } i = n , \, j=1\\
		0, & \text{otherwise }, \\
	\end{cases} \nonumber\\
	HD^{(m)}_t(i,j)  &=
	\begin{cases}
		\rho, & \text{for } i = 1 , \, j=2\\
		\bar{\rho}, & \text{for } i = T-t+1 , \,  j=1\\
		0, & \text{otherwise}.
	\end{cases} 
\end{align}
\end{lemma}
%%%%%%%%%%%%%%%%%%%%%%%%%%%%%%%%%%%%%%%%%%%%%

\subsubsection{\textbf{Temporal Analysis of the OLRA Scheme}} \label{subsubsec: OLRA analysis}

%%%%%%%%%%%%%%%%%%%%%%%%%%%%%%%%%%%%%%%%%%%%
The OLRA scheme implements proactive packet repetition with rate adaptation to improve the transmission reliability of feedback-free IoT networks. In this scheme, the packet is divided into $n$ fragments and transmitted with the rate $R_n$ several times. In particular, given that $\kappa=\lfloor \frac{T}{n}\rfloor$ and $\tau=\mod(T,n)$, each fragment is sent $\kappa$ times while the remaining $\tau$ time slots are exploited to transmit each of the randomly selected $\tau <n$ fragments one more time. In other words, the $i^{\text{th}}$ fragment is sent $\epsilon_i=\kappa+\boldsymbol{\mathbbm{1}}_{i}$, $i\in \{1,2,\cdots,n\}$ times, where $\boldsymbol{\mathbbm{1}}_{i}$ is an indicator function which is equal to $1$ if the fragment is sent $(\kappa+1)$ times of probability $\frac{\tau}{n}$ and $0$ if the fragment is sent $\kappa$ times of probability $(1-\frac{\tau}{n})$. Consequently, the test receiver has $\epsilon_i$ possible decoding chances for the $i^{\text{th}}$ fragment.

Fig.~\ref{MC for OLRA} shows the absorbing MC of the OLRA scheme for a given tuple ($n=3$, $T=11$), therefore $\tau=2$ and $\kappa =3$. For illustration, the figure assumes that fragments $\{a,c\}$ are selected to be sent $\kappa+1=4$ times while fragment $b$ is sent $\kappa=3$ times. The subscript $i$ in $a_i$ denotes the $i$-th transmission/decoding attempt of fragment $a$. As shown in Fig.~\ref{MC for OLRA}, the PSD is conditioned by the successful decoding of all the $n$ fragments. For the $i^{\text{th}}$ fragment, if all the $\epsilon_i$ decoding attempts fail, the packet is discarded, and the receiver goes into sleeping mode until the reception of the next packet. In contrast, if any decoding attempt of a fragment, except the last fragment, succeeded, the receiver has to wait until receiving the subsequent fragment and follow the same procedure. This waiting is represented by a transition into a success logic state (LS), which records the required FSD event. For the last fragment, the successful decoding of any attempts directly leads to packet absorption into the success state. Fig.~\ref{MC for OLRA} reveals that, in the OLRA scheme, a similar pattern exists for all fragments, except for the last fragment, which should be reflected in the structure of the transition matrix $\P_{\text{OLRA}}^{(m)}$ that characterizes the absorbing MC. Thus, the transition matrix $\P_{\text{OLRA}}^{(m)}$ of the OLRA scheme in Fig.~\ref{MC for OLRA} is given by \eqref{P-OLRA}, where LS in the columns and rows labels denotes the success logic state shown in Fig.~\ref{MC for OLRA}. The transition matrix $\tilde{\P}_{\text{OLRA}}^{(m)}$ in (\ref{P-OLRA}) shows similar transient and absorbing matrices for fragments $a$ and $b$ that differ from those of fragment $c$. The dashed horizontal lines separate the $\epsilon_i, \forall i$ submatrices that track the decoding trials of the $i^{\text{th}}$ fragment.
\vspace{-10pt}
\begin{figure}[H]
{\footnotesize
	\renewcommand\arraystretch{0.8}
 \makeatletter\setlength\BA@colsep{0.9pt}\makeatother
\begin{align}
&\tilde{\P}_{\text{OLRA}}^{(m)}  = \left[\begin{array}{c !{\vline width 1pt} @{\hspace{1ex}} c}
			\Q_{\text{OLRA}}^{(m)} & \H_{\text{OLRA}}^{(m)}
		\end{array} \right] = \nonumber \\[5pt]
		& \def\ra{\color{red}}
	\def\rb{\color{blue}}
	\begin{blockarray}{c @{\hspace{1cm}} c @{\hspace{1ex}} *2{c} *2{c} *2{c} {c}  *2{c} *2{c}|{c} {c} {c}|{c}{c}*2{c}}
 %%%
	\begin{block}{c @{\hspace{1cm}} c @{\hspace{1ex}} *2{c} *2{c} *2{c} {c}  *2{c} *2{c}{c} {c} {c}{c}{c}*2{c}}
	 & &  \rb {a}_2 & \rb \text{LS} &	\rb {a}_3 & \rb \text{LS} &	\rb {a}_4 &\rb \text{LS} & \rb {b}_1 &	\rb {b}_2 &\rb \text{LS} & \rb {b}_3 &	\rb \text{LS} &\rb {c}_1 & \rb {c}_2 &	\rb {c}_3 & \rb {c}_4 & & \rb \text{S} & \rb \text{t-out} \\[-3pt]
\end{block}
 %%%%%%%%%%%%%%%%%%%%%%
\begin{block}{c @{\hspace{1ex}} c @{\hspace{1ex}} [*2{c}| *2{c}| *2{c} |{c} | *2{c}| *2{c}|{c}| {c} |{c}|{c}|{c}!{\vline width 1.5pt}*2{c}]}
\ra t=1 & \rb {a}_1 & \rm \bar{p} & \rm p &  &  &  &  &  &  &  &  &  &  &  &  &  &  &  &   \bigstrut \rule{0pt}{4.5ex} \\   \cline{1-20}
% %%%%%%%%%%%%%%%%%%%%
 & \rb {a}_2 &  & & \rm \bar{p} & \rm p  &  &  &  &  &  &  &  &  &  &  &  &  &  &   \Tstrut \\
 \ra t=2 & \rb \text{LS} &  & & 0 & 1  &  &  &  &  &  &  &  &  &  &  &  &  &  &  \\	\cline{1-20}
% % %%%%%%%%%%%%%%%%%%%%
 & \rb {a}_3 &  & &  &   & \rm \bar{p} & \rm p &  &  &  &  &  &  &  &  &  &  &  &   \Tstrut \\
 \ra t=3 & \rb \text{LS} &  & &  &   & 0 & 1  &  &  &  &  &  &  &  &  &  &  &  &  \\	\cline{1-20}
% % %%%%%%%%%%%%%%%%%%%%
 & \rb {a}_4 &  & &  &   & & & \rm p  &  &  &  &  &  &  &  &  &  & 0 & \rm \bar{p}  \Tstrut \\
 \ra t=4 & \rb \text{LS} &  & &  &   &  &   & 1 &  &  &  &  &  &  &  &  &   & 0 & 0 \Tstrut \Tstrut\\	\cdashline{1-20} 
% % %%%%%%%%%%%%%%%%%%%%%%%%%
\ra t=5 & \rb {b}_1 &  & &  &   &  &   &  & \rm \bar{p} &  \rm p &  &  &  &  &  &  &  &  &  \Tstrut \\	\cline{1-20}
% % %%%%%%%%%%%%%%%%%%%%%%%%%%%%
 & \rb {b}_2 &  & &  &   &  &   &  &  &   & \rm \bar{p} & \rm p &  &  &  &  &  &  &  \Tstrut\\
 \ra t=6 & \rb \text{LS} &  & &  &   &  &   &  &  &  & 0 & 1 &  &  &  &  &  &  & \\	\cline{1-20}
% % %%%%%%%%%%%%%%%%%%%%%%%%%%%%
 & \rb {b}_3 &  & &  &   &  &   &  &  &   &  &  & \rm p &  &  &  &  & 0 & \rm \bar{p} \Tstrut\\
 \ra t=7 & \rb \text{LS} &  & &  &   &  &   &  &  &  &  &  & 1 &  &  &  &  & 0 & 0 \\	\cdashline{1-20}
% % %%%%%%%%%%%%%%%%%%%%%%%%%%
\ra t=8 & \rb {c}_1 &  & &  &   &  &   &  &  &  &  &  &  & \rm \bar{p} &  &  &  & \rm p  & 0 \Tstrut \\	\cline{1-20}
% % %%%%%%%%%%%%%%%%%%%%%%%%%%
\ra t=9 & \rb {c}_2 &  & &  &   &  &   &  &  &  &  &  &  &  & \rm \bar{p} &  &  & \rm p & 0 \Tstrut\\	\cline{1-20}
% % %%%%%%%%%%%%%%%%%%%%%%%%%%
\ra t=10 & \rb {c}_3 &  & &  &   &  &   &  &  &  &  &  &  &  &  & \rm \bar{p}  &  & \rm p & 0 \Tstrut\\	\cline{1-20}
% % %%%%%%%%%%%%%%%%%%%%%%%%%%
\ra t=11 & \rb {c}_4 &  & &  &   &  &   &  &  &  &  &  &  &  &  &   & 0 & \rm p & \rm \bar{p} \Tstrut \bigstrut \\	
\end{block}
\end{blockarray}
\label{P-OLRA}
\end{align}	}
\end{figure}
%%%%%%%%%%%%%%%%%%%%%%%%%%%%%%%%%%%%%%%%%%%%%%%
 The following lemma provides the general OLRA transition matrix  $\tilde{\P}_{\text{OLRA}}^{(m)}$ for arbitrary tuple $(n, \, T)$. 
\begin{lemma}[\textbf{OLRA Transition Matrix}] \label{Lemma: OLRA}
    \normalfont The transition matrix $\tilde{\P}_{\text{OLRA}}^{(m)}$ describing the absorbing MC for a generic packet sent from an $m$-FSD class IoT device consists of the submatrices $\Q^{(m)}_{t,\text{OLRA}}$ and $\H^{(m)}_{t,\text{OLRA}}$ for an arbitrary tuple $(n, T)$, which are given by

\begin{align} 
	\tilde{\P}_{\text{OLRA}}^{(m)}  & = \left[\begin{array}{c !{\vline width 1pt} @{\hspace{1ex}} c}
			\Q_{\text{OLRA}}^{(m)} & \H_{\text{OLRA}}^{(m)}
		\end{array} \right]\nonumber \\[5pt]
  &= \left[\begin{array}{c c c c !{\vline width 1pt} @{\hspace{1ex}} c}
  \Q_{1}^{(m)} & & &  & \H_1^{(m)}\\
         & \Q_{2}^{(m)}& & & \H_2^{(m)}\\
        &  & \ddots& & \vdots\\
        & & &\Q_{n}^{(m)} & \H_n^{(m)}
    \end{array} \right]
    \label{P_OLRA}
\end{align}
where 
\begin{equation}
	\Q_{i}^{(m)} = \begin{bmatrix}
	    \Q_{i,1} & & & \\
                 & \Q_{i,2}& & \\
                 &  & \ddots&\\
                 & & &\Q_{i,\epsilon_i}
	\end{bmatrix},\, i \in\{1,2,\cdots,n-1\}\nonumber
 \end{equation}
 \vspace{-5pt}
 \begin{align}
	 \Q_{i,1} & = \begin{bmatrix}
			\rm \bar{p} & \rm p
		\end{bmatrix},  \quad \Q_{i,2\leq k \leq \epsilon_i-1} =   \begin{bmatrix}
			\rm \bar{p} & \rm p\\
                0 &1
		\end{bmatrix}, \nonumber\\
 \Q_{i,\epsilon_i} &=  
  \begin{bmatrix}
			\rm p\\
			1
		\end{bmatrix}, \quad \Q_n^{(m)}= \rm \bar{p} \, \mathbf{I}_{\epsilon_n-1}
\end{align}
and 
\begin{equation}
\H_{i}^{(m)} = \begin{bmatrix}
    \H_{i,1}\\
    \H_{i,1}\\
    \vdots\\
    \H_{i,\epsilon_i}\\
\end{bmatrix},\, \, i \in\{1,2,\cdots,n-1\}\nonumber
\end{equation}
 \vspace{-5pt}
\begin{align}\small
     & \H_{i,1} = \o \big|_{1\times 2}, \, \,  \H_{i,2\leq k\leq \epsilon_i-1 } = \o \big|_{2 \times 2}, \, \,  \H_{i,\epsilon_i }= \begin{bmatrix}
			0 & \rm \bar{p}\\
			0 & 0 \\
		\end{bmatrix},\nonumber\\
  & \H_{n,1\leq k\leq \epsilon_n-1 } = \begin{bmatrix}
       \rm {p}& 0
   \end{bmatrix}, \quad \H_{n,\epsilon_n }= \begin{bmatrix}
			\rm {p} & \rm \bar{p}
		\end{bmatrix},
\end{align}
\end{lemma}
 \vspace{-10pt}
\begin{remark}\normalfont
For the sake of energy saving, another variant of the OLRA scheme is suggested. In this scheme, each fragment is sent $\kappa=\lfloor T/n\rfloor$ times, and the remaining $\tau=\mod{(T,n)}$ time slots are kept silent. We denote this scheme as~\textbf{\emph{OLRA with energy saving (OLRA-ES)}}. The transition matrix $\tilde{\P}_{\text{OLRA-ES}}^{(m)}$ of the OLRA-ES scheme is similar to the transition matrix of the OLRA scheme given in~\eqref{P_OLRA} with $\epsilon_i=\kappa$ instead of $\kappa+\boldsymbol{\mathbbm{1}}_{i}$ for fragments $i \in \{1,2,\cdots,n\}$. 
\end{remark}
%%%%%%%%%%%%%%%%%%%%%%%%%%%%%%%%%%%%%%%%%%%%%%
 \vspace{-10pt}
\subsection{Spatial Analysis}\label{spatial analysis}
In this subsection, the network-wide spatial analysis is conducted using stochastic geometry to characterize the FSD probability for a generic $m$-FSD device $\rm p_{n,m}$ and the feedback success probability $\rm p_{ack}$ that controls the transition matrices given in Lemmas~\ref{Lemma: CLRA} and~\ref{Lemma: OLRA}. 
 \vspace{0.2cm}
\subsubsection{\textbf{Forward Fragment Success Delivery Probability}} 
Thanks to the independent thinning property of the PPP, the set of IoT devices of type $v \in \V$ is an independent PPP denoted by $\Psi_v \subset \Psi$ with intensity $\lambda_v= f_{\rm v}(v) \lambda$. The FSD probability is defined as the probability that the received signal at the test receiver located at the origin is larger than the SIR detection threshold $\theta_n$ when the packet is transmitted with the rate $R_{n}$. Therefore, $\rm p_n$ can be expressed as
\begin{align}
\rm p_n  &= \mathbb{P} \left \{\text{SIR} \geq \theta_n \big| \Psi \right\} \nonumber \\
& = \mathbb{P} \left \{\frac{p_o h_o ||\rm x_o||^{-\eta}}{\sum_{v=1}^{V}\sum_{\rm x_i \in \Psi_v \setminus \rm x_o} \boldsymbol{\mathbbm{1}}_{\xi_i} \, p_v h_i ||\rm x_i|| ^{-\eta}} \geq \theta_n \bigg|\Psi \right\},
 \label{SDP}
\end{align}
where $p_o$ is the power of the intended signal, $h_o$ is the desired channel gain, $||\rm x_o||$ is the distance between the test receiver at the origin and its test transmitter, $\rm x_i \in \Psi_v$ is the location of an interfering IoT device of type $v$, $h_i$ is the interfering channel gain, $\boldsymbol{\mathbbm{1}}_{\xi_i}$ is an indicator function which is equal to one if the $\rm x_i$ interfering IoT device is active and equal zero otherwise, and $||\rm x_i||$ is the distance between the $i$-th interfering device and the test receiver. For an arbitrary realization of the HPF, the FSD $\rm p_n$ is a function of the locations of the interfering devices. Such realization-dependent FSD is fully characterized via the meta distribution of the FSD probability, which is defined as
\begin{equation}
	\bar{F}_s(\theta_n,\delta)=\mathbb{P}\left\{\mathbb{P} \left\{ \text{SIR}\geq \theta_n \big|\Psi \right\} > \delta \right\}  = \mathbb{P} \left\{\rm p_n > \delta \right\}.
	\label{meta_def}
\end{equation}
$\bar{F}_s(\theta_n,\delta)$ defines the likelihood that the decoding process at the test receiver within an arbitrary realization of the HPF, with a detection threshold $\theta_n$, succeeds for more than $\delta$ percent of the time. Thus, for a given set of parameters $n$ and $\delta$, (\ref{meta_def}) generalizes the FSD model for all realizations of the HPF. For analytical tractability, the meta distribution in (\ref{meta_def}) is approximated using the beta approximation as follows~\cite{haenggi2015meta} 
% {\small
% \begin{equation}
% 	\bar{F}_s(\theta_n,\delta) \approx  1- \mathcal{I}_{\delta} \left( \! \frac{M_1 (M_1-M_2)}{M_2-M_1^2}, \frac{(1-M_1)(M_1-M_2)}{M_2-M_1^2} \!\right),
% 	\label{meta_approx}
% \end{equation}}
\begin{equation}
	\bar{F}_s(\theta_n,\delta) \approx  1- \mathcal{I}_{\delta} \left( M_1 \mathcal{X}, (1-M_1) \mathcal{X}\right),
	\label{meta_approx}
\end{equation}
where $M_1$ and $M_2$ are the first two moments of the FSD probability at
rate $R_n$, $\mathcal{X}=(M_1-M_2)/(M_2-M_1^2)$, $\mathcal{I}_{\delta}(a, b) = \frac{1}{\mathcal{B}(a,b)} \int_{0}^{\delta} t^{a-1}(1 - t)^{b-1}dt$ is the regularized incomplete beta function, and $\mathcal{B}(x,y)= \int_{0}^{\infty} \frac{t^{x-1}}{(1+t)^{x+y}} \text{d}t $ is the beta function. The moments $M_1$ and
$M_2$ of the FSD probability are given as in the following lemma.
\begin{lemma}[Moments of FSD probability] \label{Lemma: moments of meta distribution} \normalfont The moments $M_1$ and $M_2$ of the FSD probability $\rm p_n$ are given as
\begin{equation}
	M_1  = \exp \left[\frac{- 2 \pi^2  R_o ^2 \, \theta_n^{\frac{2}{\eta} }}{\eta \sin \left(\frac{2 \pi}{\eta} \right)} \sum_{v=1}^{V} \left(\frac{p_v}{p_o}\right)^{\! \! \frac{2}{\eta}}\lambda_v \alpha_v \right], 
 \label{1st_moment}
 \end{equation}

 {\small
 \begin{equation}
M_2 =\!\exp \Bigg[\frac{- 2 \pi^2  R_o ^2 \, \theta_n^{\frac{2}{\eta} } }{\eta \sin \left(\frac{2 \pi}{\eta} \right)} \sum_{v=1}^{V} \!  \left(\frac{p_v}{p_o}\right)^{\! \! \frac{2}{\eta}}\lambda_v \alpha_v \! \left(2-\alpha_v \!  \left(\! 1-\frac{2}{\eta}\right)\right)\!\Bigg]. 
 \label{2nd_moment} 
\end{equation}}
	
 \begin{IEEEproof}
%By averaging the FSD probability $\rm p_n$ in (\ref{SDP}) over the interfering devices types, activities, locations, and the fading gains, and after some manipulations of the integrals, the moments $M_1$ and $M_2$ in (\ref{1st_moment}) and (\ref{2nd_moment}) are obtained. For detailed proof, please refer to Appendix A.
See Appendix~\ref{app:lemma4}. 
\end{IEEEproof}
\end{lemma}

To conduct the queuing analysis, the approximated meta distribution in (\ref{meta_approx}) is discretized into $M$ equiprobable FSD classes. To define the FSD probability 
ranges for the different classes, we set $w_0 = 0$ and $w_M = 1$,
and define the set $\{w_2, w_3, \cdots, w_{M-1}\}$ such that
\begin{equation}
	\bar{F}_s(\theta_n, w_m) - \bar{F}_s(\theta_n, w_{m-1})= \frac{1}{M}.
 \label{meta_discretized}
\end{equation}
The FSD probabilities within the range $[w_m, w_{m+1}]$ are approximated via
the median value $\rm p_{n,m}$, which is given by
\begin{equation}
	\bar{F}_s(\theta_n, w_m) - \bar{F}_s(\theta_n,\mathrm{p_{n,m}})= \frac{1}{2 M}.
	\label{discretize2}
\end{equation}
The above discretization implies that the likelihood for the intended IoT link to operate with any of the FSD probabilities $\rm p_{n,m}$ is $\frac{1}{M}$. Using the discretized $\rm p_{n,m}$, we can build a queueing model for each FSD class for the CLRA and OLRA schemes as shown in Section~\ref{subsec: temporal analysis}. 
\subsubsection{\textbf{Feedback Success Probability}}
%As aforementioned, the PSD probability of the CLRA scheme depends on the tuple $(\rm p_{n,m} \, \forall m, \rm p_{ack})$. This section provides an expression of the feedback success probability $\rm p_{ack}$, which 
The feedback success probability $\rm p_{ack}$ is defined as the probability that the feedback SIR is larger than a detection threshold $\theta_{\text{ack}}$. Assuming that the typical transmitter $\rm x_o$, located at the origin, receives a feedback message (i.e., either ACK or NACK) from its corresponding receiver $\rm y_o$, the feedback success probability is given as 
\begin{align}
    \rm p_{ack} &= \mathbb{P} \left\{\text{SIR}_{\text{ack}} \geq \theta_{\text{ack}}\right\}\nonumber\\
    &= \mathbb{P} \left\{\frac{p_t h_o ||\rm y_o-x_o||^{-\eta}}{\sum_{\rm y_i \in \Psi \setminus \rm y_o } p_t h_i ||\rm y_i-x_o||^{-\eta}} \geq \theta_{\text{ack}} \right\}
    \label{feedback success prob}
\end{align} 
where $p_t$ is the feedback transmit power. We assume that all interfering IoT receivers $\rm y_i \in \Psi$ have the same feedback transmit power $p_t$. The feedback success probability $\rm p_{ack}$ is given by the following Lemma.
\begin{lemma} [Feedback success probability]\normalfont
    The feedback success probability defined in~\eqref{feedback success prob} is given as  
    \begin{equation}
        \rm {p_{ack}} =\exp\left[\frac{-2 \pi^2 \lambda R_o^2 \, \theta_{{ack}}^{\frac{2}{\eta}}}{\eta \sin \left(\frac{2\pi}{\eta}\right)}\right]
        \label{eq:pack}
    \end{equation}
    \begin{IEEEproof}
    The proof follows a similar approach as in Lemma~\ref{Lemma: moments of meta distribution}, thus omitted. 
    % By averaging the feedback success probability $\rm p_{ack}$ given in~\eqref{feedback success prob} over intended feedback and interfering channel gains and the spatial locations of IoT devices, $\rm p_{ack}$  acn be expressed as
    % \begin{align}
    %     \rm {p_{ack}} &= \mathbb{E}_{h_o,h_i,\Psi} \left\{h_o \geq \theta_{\text{ack}} R_o^\eta {\sum_{\rm y_i \in \Psi \setminus \rm y_o } R_i^{-\eta} h_i} \right\} \nonumber\\
    %     &= \mathbb{E}_\Psi\left\{\prod_{\rm y_i \in \Psi \setminus \rm y_o } \frac{1}{1+\theta_{\text{ack}} \left(\frac{R_o}{R_i}\right)^{\eta}}\right\} \nonumber\\
    %     &= \exp \left[-2\pi\lambda \int_{0}^{\infty} \left(1-\frac{1}{1+\theta_{\text{ack}} \left(\frac{R_o}{v}\right)^\eta} \right) v dv\right]\nonumber\\
    %     &= \exp \left[\frac{-2\pi\lambda R_o^2 \, \theta_{\text{ack}}^{2/\eta}}{\eta} \int_{0}^{\infty} \frac{y^{\frac{-2}{\eta}}}{1+y}dy\right]\nonumber\\
    %     &=\exp\left[\frac{-2 \pi^2 \lambda R_o^2 \theta_{\text{ack}}^{\frac{2}{\eta}}}{\eta \sin \left(\frac{2\pi}{\eta}\right)}\right]
    % \end{align}
    \end{IEEEproof}
    \label{Lemma: feedback success probability}
\end{lemma} 
%%%%%%%%%%%%%%%%%%%%%%%%%%%%%%%%%%%%%%%%%%%%%%%%%%%%%%%%%
\subsection{Performance Metric Analysis} \label{prerformance metric analysis}
In this section, the main performance metrics of the CLRA and OLRA schemes are formulated given the transition matrices provided in Lemmas~\ref{Lemma: CLRA} and~\ref{Lemma: OLRA}. By referring to the MAM, the PSD probability and mean latency for a device in the $m$-th FSD class are expressed in  Theorem~\ref{theorem 1}.
\begin{theorem} \label{theorem 1} \normalfont Let the vector $\A^{(m)} \in \{A_s^{(m)}, A_f^{(m)} \} $ defines the probability that a generic packet of a device that belongs to the $m$-th FSD class is eventually absorbed into the success or timeout (failure) states, respectively, where $A_f^{(m)}= 1- A_s^{(m)}$. In addition, let $\D^{(m)} \in \{D_s^{(m)}, D_f^{(m)}\}$ denotes a scaled mean latency (delay) to absorption for a device in the $m$-th FSD class, where $\frac{D_s^{(m)}}{A_s^{(m)}}$ and $\frac{D_f^{(m)}}{A_f^{(m)}}$ are, respectively, the packet mean latency, in time slots, for absorption into success and timeout states. Thus, $\A^{(m)}$ and $\D^{(m)}$ can be defined as~\cite{alfa2016applied} 
\begin{equation}\small
\A^{(m)}  = \H_1^{(m)}+\sum_{i=2}^{T-2} \left(\prod_{t=1}^{i-1} \Q_t^{(m)} \right) \times H_i^{(m)}. \label{success_prob} \end{equation} 
\begin{equation}\small
\D^{(m)}  = \H_1^{(m)}+\sum_{i=2}^{T-2} \left(\prod_{t=1}^{i-1} i \Q_t^{(m)} \right) \times H_i^{(m)}. \label{success_latency}
\end{equation}
\begin{IEEEproof}
According to~(\ref{transition matrix}), the transition probability within the decoding attempts before the absorption into a final state is captured by the transient matrix $\Q^{(m)}$ and the probability of absorption is captured by $\H^{(m)}$. Hence, the probability that the receiver handles $t$ consecutive decoding attempts is given by $\prod_{i=1}^{t}\Q_i^{(m)}$. Similarly, the probability that a generic packet is absorbed after exactly $t$ time slots is given by $\prod_{i=1}^{t-1} \Q_i^{(m)} \H_t^{(m)}$. Therefore, by applying the law of total probability and accounting for the $T$- time slot packet deadline, Theorem \ref{theorem 1} is proved.
\end{IEEEproof}
\end{theorem}
Note that $\D^{(m)}$ in~\eqref{success_latency} determines the average number of time slots for packet absorption into success or timeout states. The mean packet round-trip latency depends on the adopted transmission strategy. Therefore, the mean packet latency for the CLRA and OLRA schemes is given in Corollary~\ref{corollary:latency}. 
\begin{corollary} \label{corollary:latency}
\normalfont 
The mean absorption (to transmission success or elapsed deadline) latency of a generic packet for a device belonging to the $m$-FSD class for the CLRA scheme is given as 
\begin{equation}
   \D_{\text{CLRA}}^{(m)} = \D^{(m)} (T_s+T_{\text{ack}}) \quad \text{sec}, \end{equation}
and for the OLRA scheme, the mean absorption latency is given as
   \begin{equation}
   \D_{\text{OLRA}}^{(m)} = \D^{(m)} T_s \quad \text{sec},
\end{equation}
where $T_{\text{ack}}$ is the feedback message duration, which is significantly shorter than the fragment duration $T_s$ owing to the short ACK/NACK messages.

% \begin{IEEEproof}
% According to the adopted policy of CLRA scheme, a generic fragment's round-trip latency includes the transmission/decoding process, that is assumed to be fit within a time slot $T_s$, and feedback ACK/NACK sent from test receiver back to the transmitter of $T_{\text{ack}}$ duration. On the other side, the feedback is absent in the OLRA transmission schemes (i.e., OLRA-FR and OLRA-VR).
% \end{IEEEproof}
\label{corollary1}
\end{corollary} 

The overall energy consumption of a generic test receiver that belongs to the $m$-FSD class depends on the adopted transmission policy. In the OLRA scheme, the test receiver consumes energy in packet fragment reception and decoding. In the CLRA scheme, the receiver consumes extra energy in feedback ACK/NACK signaling transmission. The energy consumptions of the test receiver in the fragment reception and decoding and the feedback acknowledgment denoted as $E_r$ and $E_{\text{ack}}$, are given as~\cite{amin2012cooperative,al2019bound}
\begin{equation}
\begin{aligned}
    E_r &= p_{cr}  T_s,\\
    E_{\text{ack}} &= \left(\gamma  p_t + p_{ct}\right) T_{\text{ack}},
\end{aligned}
\end{equation}
where $p_{cr}$ is the radio frequency (RF) circuit power consumption at the receiver, $p_{ct}$ denotes the RF circuit power consumption of the test receiver's transceiver in transmitting the ACK/NACK feedback messages, $p_t$ is the feedback transmitted power, and $\gamma$ is the power amplifier conversion factor of value at least $1$. 

Finally, the total energy consumption during packet delivery depends on the mean latency for packet absorption, whether to success or time-out states. The total energy consumption of a receiver belonging to the $m$-FSD class is given by the following corollary. 
\begin{corollary} 
\normalfont 
The overall energy consumption of an $m$-FSD class device for the CLRA and OLRA transmission schemes is given by
\begin{equation}\begin{aligned}
    \E_{\text{CLRA}}^{(m)} &= (E_r +E_{\text{ack}})\left(D_{s,\text{CLRA}}^{(m)}+D_{f,\text{CLRA}}^{(m)} \right),\nonumber\\
   \E_{\text{OLRA}}^{(m)} &= E_r \left(D_{s,\text{OLRA}}^{(m)}+D_{f,\text{OLRA}}^{(m)} \right), 
\end{aligned}
   \label{energy_consumption}
\end{equation}

% \begin{IEEEproof}
% {\color{blue}According to the OLRA transmission scheme, the test receiver consumes energy in packet fragment reception and decoding. For the CLRA scheme, the feedback ACK/NACK signaling consumes extra energy. Therefore, the energy consumption of the OLRA scheme in a fragment delivery is $E_r$ while for CLRA is $(E_r+E_{\text{ack}})$. Consequently, the overall energy consumption is obtained by multiplying the energy consumption in a single attempt by the weighted number of time slots the receiver spends in packet delivery success/failure. Applying the law of total probability, the time the receiver spends for a packet delivery is the sum of PSD latency $\frac{D_s^{(m)}}{A_s^{(m)}}$ times the PSD probability $A_s^{(m)}$ and packet delivery failure latency $\frac{D_f^{(m)}}{A_f^{(m)}}$ multiplied by the probability $A_f^{(m)}$ that the packet deadline is elapsed before the PSD.}
% \end{IEEEproof}
\label{corollary2}
\end{corollary}
% \vspace{-5pt}
Finally, by averaging over the FSD classes, the probability that a packet is absorbed to either success or failure states, $\A$, the mean latency for such absorption $\D$, and the energy consumption at the test IoT receiver are given as 
\begin{align}
  \A &=\mathbb{E}_m \left\{\A^{(m)}\right\}\nonumber\\
  \D_x &=\mathbb{E}_m \left\{\D_x^{(m)} \right\} \nonumber\\
  \E_x &=\mathbb{E}_m \left\{{\E_x}^{(m)}\right\}
\end{align}
where $ x \in \{\text{CLRA},\text{OLRA}\}$.
%%%%%%%%%%%%%%%%%%%%%%%%%%%%%%%%%%%%%%%%%%%%%%%%%%%%
% \vspace{-20pt}
\section{Numerical Results} \label{sec: numerical results}
In this section, we provide numerical results to highlight the impact of the repetition mechanism and the feedback presence/absence on the developed CLRA and OLRA schemes. Monte Carlo simulations are carried out to validate the accuracy of the developed analytical models. Unless otherwise stated, the simulation parameters for the network are provided in Table~\ref{table: Simulation parameters}. The Monte Carlo simulations construct the meta distribution $\bar{F}_s(\theta_n, \delta)$ across $5000$ different HPF realizations with $10^5$ time iterations per each HPF realization. For the analysis, the feedback success probability $\rm p_{ack}$ is obtained in~\eqref{eq:pack}. The meta distribution of the FSD probability for each rate $R_n$ is obtained as in~(\ref{meta_approx}). The transition matrices of the CLRA and OLRA schemes defined, respectively, in Lemmas~\ref{Lemma: CLRA} and~\ref{Lemma: OLRA} are constructed upon obtaining the FSD probability $\rm p_{n,m}, m=\{1,2,\cdots,M\}$ from~(\ref{discretize2}), which discretizes the FSD meta distribution in~(\ref{meta_approx}). By relying on the constructed matrices, the performance metrics of a generic IoT device (i.e., PSD probability, PSD mean latency and overall energy consumption) are obtained in Theorem~\ref{theorem 1}, Corollary~\ref{corollary1}, and Corollary~\ref{corollary2}.
%%%%%%%%%%%%%%%%%%%%%%%%%%%%%%%%%%%%%%%%%%%%%%%%%%%
\begin{table}[t!] 
\small \caption{\small Network and simulation parameters} % title of Table
\centering % used for centering table
\resizebox{\columnwidth}{!}{%
% \begin{adjustbox}{width=\columnwidth, height=3cm}
\begin{tabular}{c c c}
\hline 
\textbf{Symbol} & \textbf{Definition} & \textbf{Value} \\
\hline \\[-0.9ex]
$\lambda$ & Intensity of PPP $\Psi$ & $2 \times 10^2$ devices/Km$^2$\\ 
$\eta$ & Fading exponent & $4$\\
$R_o$ & Distance of Tx/Rx link & $20$ meters\\
\rm V   & Types of IoT interfering devices &  $3$\\
$f_{\rm v(v)}$ & PMF of types of IoT interfering devices  & Uniform \\
$\alpha_v$ & Activity of type $v$ IoT interfering devices & $\{0.1,0.3,0.5\}$\\
$p_v$ & Transmitted power of type $v$ IoT interfering devices & $\{10,7,5\}$ mWatts\\
$p_t$ & Transmitted power of the test IoT device & $10$ mWatt\\
$W$ & Bandwidth & $250$ KHz\\
$L$ & Packet length & $300$ Bytes\\
$T_s$ & Packet duration & $1$ ms\\
$T$ & Packet deadline & $ 15 $ time slots\\
$p_{ct}$ & Tx mode RF circuit power consumption & $38$ mWatt ~\cite{semiconductor2009nrf24l01}\\
$p_{cr}$ & Rx mode RF circuit power consumption & $45$ mWatt~\cite{semiconductor2009nrf24l01} \\
$\gamma $ & Power amplifier conversion factor & $4$~\cite{al2019bound}\\
$L_{\text{ack}}$ & ACK/NACK messages length & $\{5, 15\}$ Bytes\\
$T_{\text{ack}}$ & ACK/NACK messages duration & $0.15 \, T_s = 0.15$ ms\\
\hline %inserts single line
\end{tabular}
% \end{adjustbox}
}
\label{table: Simulation parameters} 
\vspace{-0.3cm}
\end{table}
%%%%%%%%%%%%%%%%%%%%%%%%%%%%%%%%%%%%%%%%%%%%%%%%%%%%%%%%%%%%%%%
\begin{figure}[t!]
	\centering
	\includegraphics[width=0.9\linewidth,height=5cm]{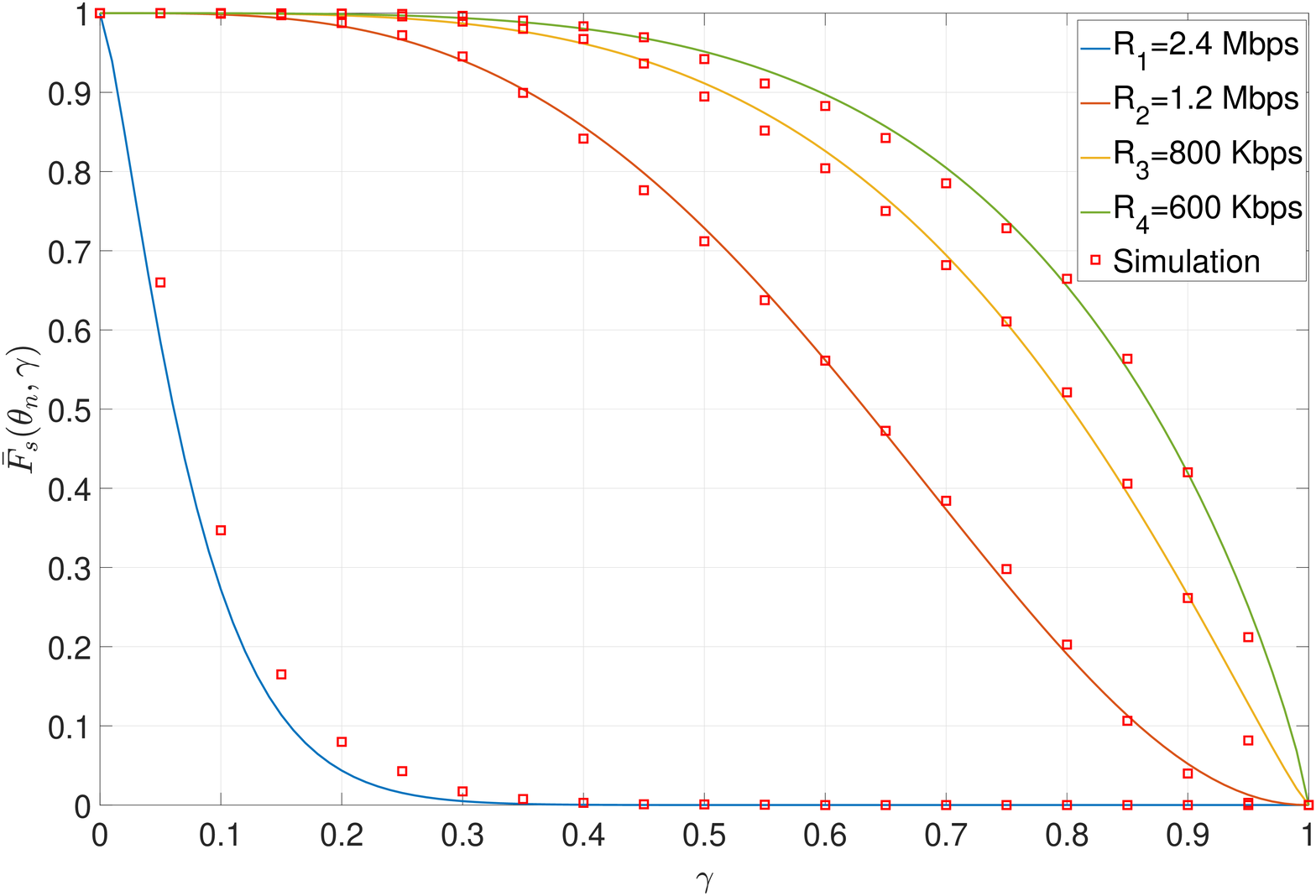}
	\small \caption{Meta distribution of the FSD at different transmission rates $R_n,\, n\in\{1,\cdots,4\}$.}
	\label{fig:meta}
 \vspace{-0.5cm}
\end{figure}

Fig.~\ref{fig:meta} shows the meta distribution of the FSD for different transmission rates. The close match between the analysis (i.e., curves) and simulations (marks) validates the beta distribution approximation of the meta distribution used in (\ref{meta_approx}).
The figure also shows the impact of transmission rate $R_n$ on the FSD probability (i.e., transmission reliability). Dividing the packet into more fragments enables a lower transmission rate and leads to a lower detection threshold $\theta_n$ that is more likely to be satisfied, which improves the transmission reliability. For instance, dividing the packet into $2$ fragments significantly improves the probability that the SIR decoding threshold $\theta_n$ is satisfied by the intended link for $20 \%$ percent of the time from $0.04$ to $0.98$. 
%%%%%%%%%%%%%%%%%%%%%%%%%%%%%%%%
\begin{figure}[t!]
	\centering
		\begin{subfigure}[t]{0.50\textwidth}
			\centering
			\includegraphics[width=\linewidth,height=5.5cm]{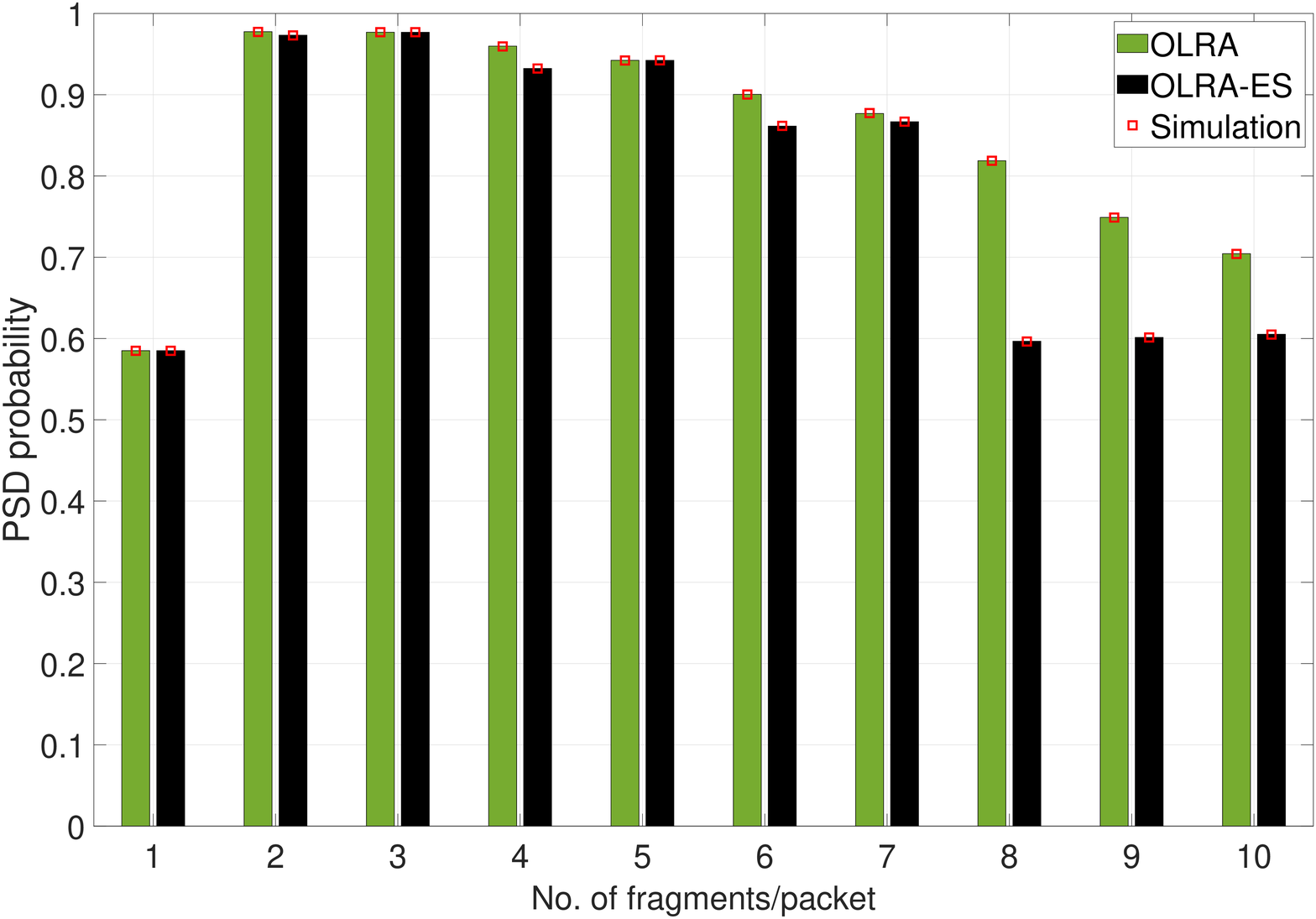}
                \small \caption{}
			% \small \caption{PSD probability of CLRA scheme.}
         \label{fig:ps_OLRA}
		\end{subfigure}%
  
		\begin{subfigure}[t]{0.50\textwidth}
			\centering
			\includegraphics[width=\linewidth,height=5.5cm]{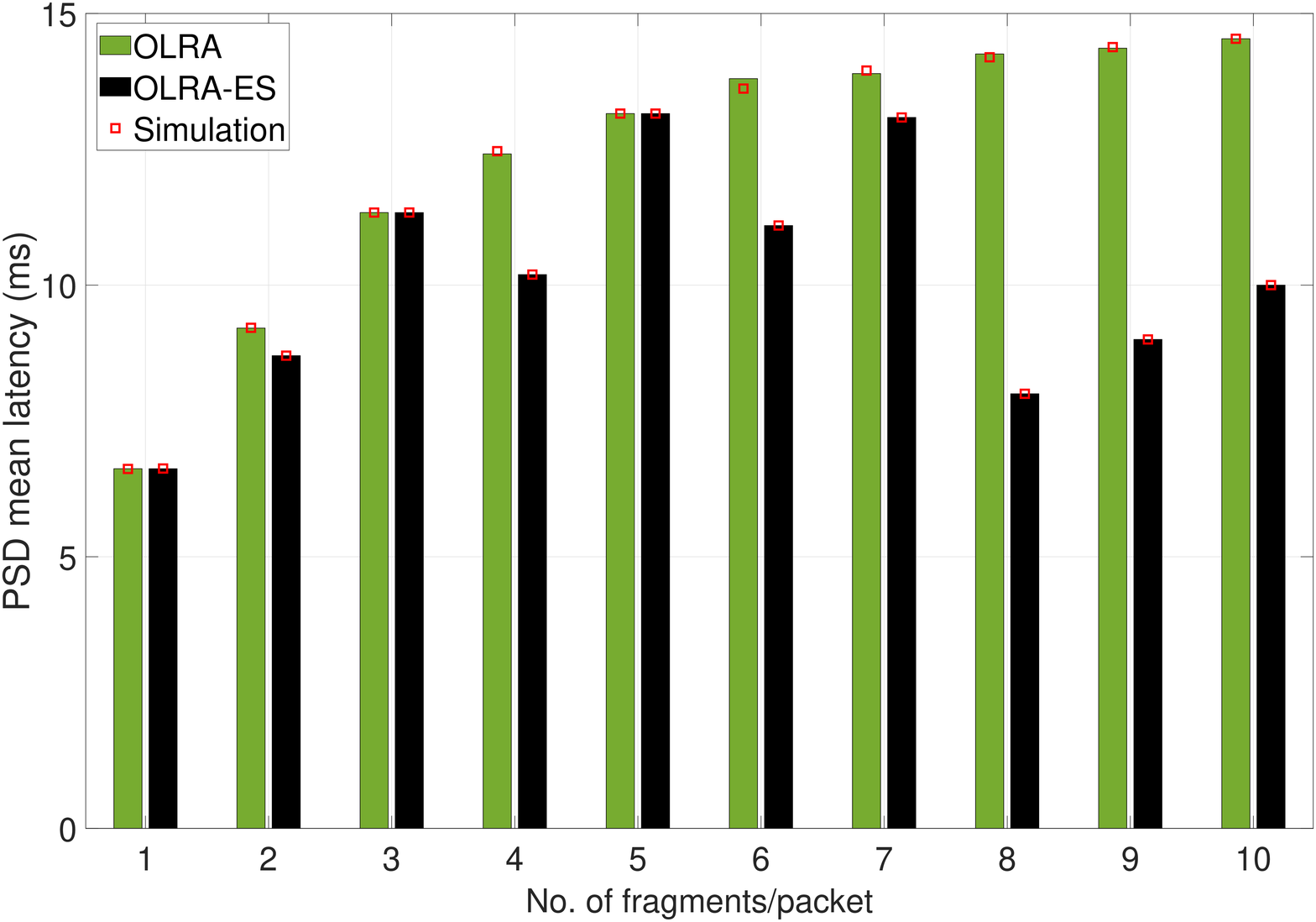}
			 \small \caption{}
         \label{fig:latency_OLRA}
		\end{subfigure}%
		\small \caption{PSD probability and mean latency for OLRA transmission schemes.}
	\label{fig:OLRA}
 \vspace{-0.6cm}
\end{figure} 
%%%%%%%%%%%%%%%%%%%%%%%%%%%%%%%%

Before comparing the CLRA and OLRA schemes to quantify the impact of feedback, we first contrast the two OLRA variants in Fig.~\ref{fig:OLRA}. In particular, Fig.~\ref{fig:OLRA} depicts the PSD probability and mean latency of the proposed OLRA and OLRA-ES transmission schemes. At first, we highlight the match between the analysis and simulation results, which validates our developed mathematical framework. Recalling from Section~\ref{subsubsec: OLRA analysis}, both schemes have a minimum repetition of $\kappa=\lfloor T/n \rfloor $ times for each fragment. However, both schemes differ in the way they exploit the remaining $\tau= \mod(T,n)$ time slots. Fig.~\ref{fig:OLRA} confirms that both schemes have similar performance when $T$ is divisible by $n$ (i.e., the number of remaining slots $\tau$ is $0$) shown for the cases of $n\in\{1,3,5\}$. For other values of $n$, the figure illustrates that OLRA outperforms OLRA-ES in terms of transmission reliability, at the expense of increased transmission latency, as depicted in Fig.~\ref{fig:latency_OLRA}. This disparity arises from the transmission of $\tau$ fragments from the $n$ fragments $\kappa+1$ times in the OLRA scheme, compared to the $\kappa$ times in the OLRA-ES scheme. The superiority of OLRA over OLRA-ES in transmission reliability is governed by the values of $\kappa$ and $\tau$. A higher value of $\kappa$ increases the likelihood of successful fragment delivery within $\kappa$ decoding trials, diminishing the significance of sending $\tau$ fragments an additional time, resulting in a less pronounced improvement in PSD probability. For instance, for $n=4$ (i.e., $\kappa =3$ and $\tau=3$), the PSD probability of OLRA surpasses that of the OLRA-ES scheme by $3\%$, with a $22.8\%$ increase in latency. Conversely, a smaller $\kappa$ signifies the impact of $\tau$ on the transmission reliability at the expense of latency increment. For instance, with $n=8$ (i.e., $\kappa=1$ and $\tau=7$), the PSD probability escalates by $37.2\%$, at the cost of a $78.3\%$ latency increment. 

\begin{figure}[t!]
	\centering
        \includegraphics[width=\linewidth,height=5.5cm]{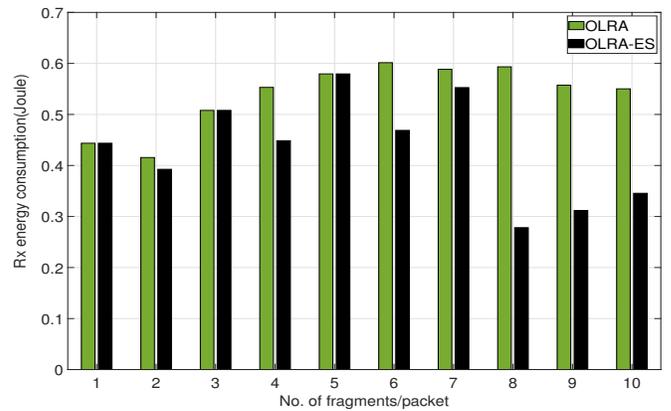}

 \small \caption{Receiver energy consumption for OLRA schemes.}
	\label{fig:Ec_OLRA}
  \vspace{-0.3cm}
\end{figure}

Fig.~\ref{fig:Ec_OLRA} illustrates the energy consumption of the test receiver when decoding a standard packet from the corresponding transmitter in both the OLRA and OLRA-ES schemes. We can clearly see that the OLRA consumes more energy due to its increased PSD latency compared to the OLRA-ES scheme, as shown in Fig.~\ref{fig:latency_OLRA}. This observation aligns with the relationship described in~\eqref{energy_consumption}.
% Fig.~\ref{fig:OLRA} plots the analytical and simulation results of the PSD probability and mean latency of the OLRA schemes (i.e., OLRA-FR and OLRA-VR) as function of the number of fragments. %Fig.~\ref{fig:ps_CLRA} shows the probability of P, while Fig.~\ref{fig:ds_CLRA} reveals the mean latency for packet successful delivery. 
% The matching between the simulation and analysis concretely validates the developed transition matrix in Lemmas~\ref{Lemma: OLRA-FR} and~\ref{Lemma: OLRA-VR}. Fig.~\ref{fig:ps_CLRA} reveals an optimal fragmentation rate that optimizes the PSD probability (i.e., $R_{4}$). Increasing further the fragmentation rate deteriorates the transmission reliability as the device will no longer be able to transmit all the fragments successfully within the packet deadline $T$. Fig.~\ref{fig:ds_CLRA} shows the cost of packet fragmentation as dividing the packet into more fragments expands the packet transmission over more time slots leading to high PSD mean latency. 
%%%%%%%%%%%%%%%%%%%%%%%%%%%%%%%%%%%
\begin{figure}[t!]
	\centering
		\begin{subfigure}[t]{0.50\textwidth}
			\centering
			\includegraphics[width=\linewidth,height=5.5cm]{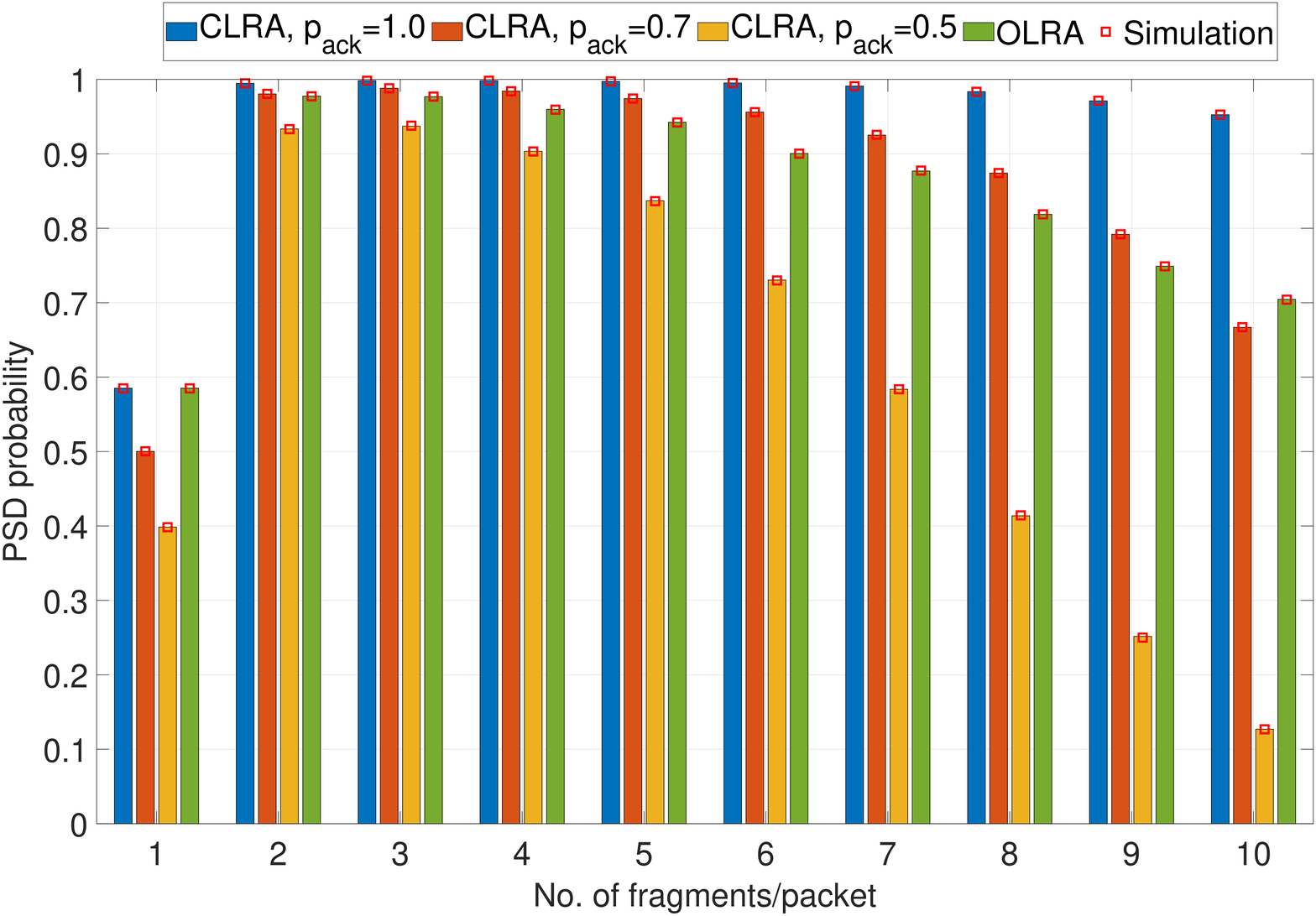}
			\small \caption{}
         \label{fig:ps_comp}
		\end{subfigure}%
		
		\begin{subfigure}[t]{0.50\textwidth}
			\centering
			\includegraphics[width=\linewidth,height=5.5cm]{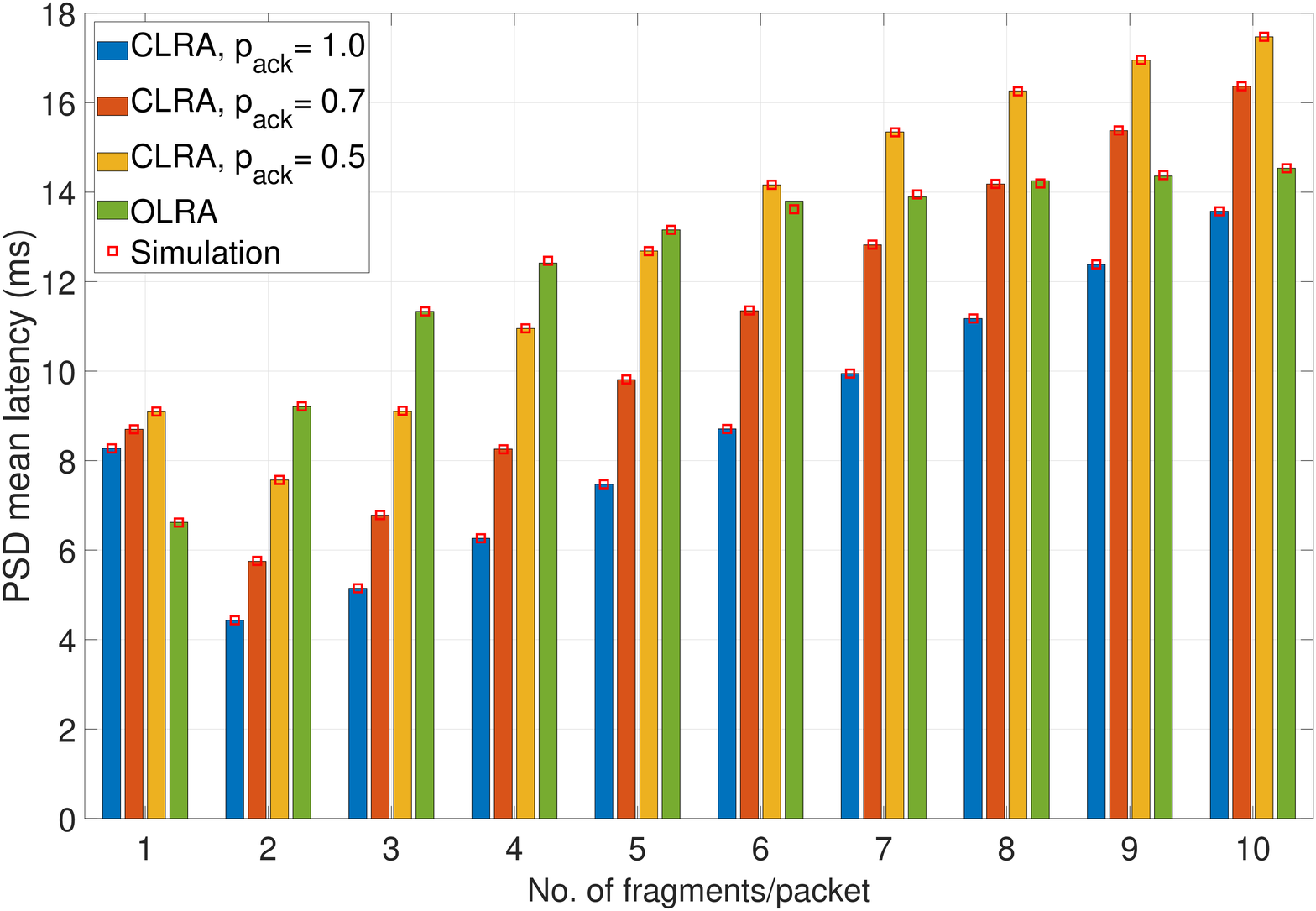}
			 \small \caption{}
         \label{fig:latency_comp}
		\end{subfigure}%
		% \small \caption{The PSD probability and mean latency for proposed OLRA transmission schemes.}
		\small \caption{Comparison between OLRA and CLRA schemes in terms of PSD probability and mean latency.}
  \label{fig:comparison}
  \vspace{-0.6cm}
\end{figure} 
%%%%%%%%%%%%%%%%%%%%%%%%%%%%%%%%%
Since the OLRA is superior to its OLRA-ES counterpart, we use the former to quantify the impact of the feedback by benchmarking against the CLRA scheme. In this context, Fig.~\ref{fig:comparison} compares the OLRA with the feedback-aware CLRA scheme for three values of $\rm p_{ack}$. The $\rm p_{ack}=1$ represents the idealistic error-free feedback channel that shows the best performance in terms of reliability and latency. Accounting for the feedback errors reveals the true performance, which can be significantly less than the idealistic case as shown for $\rm p_{ack}=0.7$ (i.e., $L_{\text{ack}}=5$ Bytes) and $\rm p_{ack}=0.5$ (i.e., $L_{\text{ack}}=15$ Bytes). The impact of the feedback is quantified by comparing the feedback-free OLRA to the CLRA scheme. It can be shown that the CLRA scheme outperforms OLRA by $2.5 \%$ when the number of fragments $n=4$ and $\rm p_{ack}=0.7$. The figure also shows that the OLRA scheme has a higher transmission latency. However, for higher feedback errors, the OLRA-FR scheme performs better in terms of reliability and latency. In all schemes, the positive impact of packet fragmentation is underscored showing an optimal number of fragments that maximizes the reliability. The PSD probability for $n>1$ (i.e., with fragmentation) is remarkably higher than for $n=1$ (i.e., without fragmentation) at the expense of increased transmission latency.

\begin{figure}[t!]
	\centering
	\includegraphics[width=\linewidth,height=5.5cm]{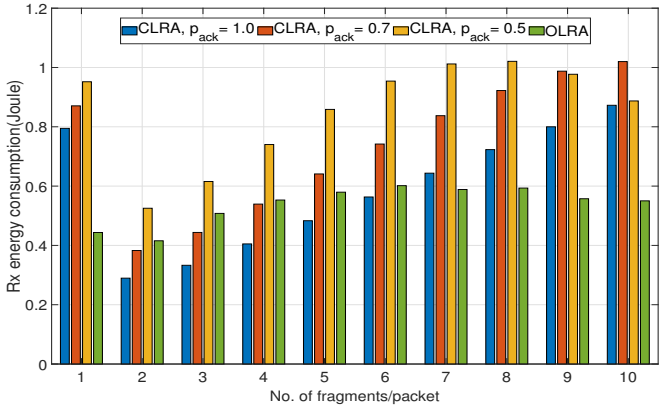}
	\small \caption{Comparison between OLRA and CLRA schemes in terms of receiver energy consumption in packet decoding.}
	\label{fig:Ec_comp}
  \vspace{-0.7cm}
\end{figure}
%%%%%%%%%%%%%%%%%%%%%%%%%%%%%%%%%%%%

Although the CLRA scheme can improve transmission reliability, it has a direct impact on the energy consumption of IoT devices. Fig.~\ref{fig:Ec_comp} highlights the cost of the feedback-aware CLRA scheme over the feedback-free OLRA scheme in terms of the overall energy consumption by the test receiver in packet reception and decoding and feedback ACK/NACK transmissions. It can be clearly shown that the receiver adopting the CLRA transmission scheme consumes more energy compared with the OLRA scheme owing to the energy consumed in the feedback transmission. Furthermore, Fig.~\ref{fig:Ec_comp} shows the negative impact of feedback channel imperfection on the receiver energy consumption for the CLRA scheme.
%%%%%%%%%%%%%%%%%%%%%%%%%%%%%%%
% \vspace{-0.5cm}
\section{Conclusion}\label{sec: conclusion}
In this paper, feedback-aware CLRA and feedback-free OLRA transmission schemes tailored for delay-sensitive and energy-constrained large-scale IoT networks are addressed. Novel and tractable spatiotemporal frameworks based on stochastic geometry and queuing theory are developed for the OLRA transmission schemes and benchmarked against the CLRA scheme under an error-prone feedback setting. The simulation outcomes validate the analytical models for both OLRA and CLRA and emphasize a crucial insight that rate adaptation substantially improves transmission reliability, at the expense of latency. Notably, our findings underscore the adverse effects of imperfections within the feedback channel on the CLRA scheme, manifested in reduced reliability and increased energy consumption. Furthermore, our results quantify the energy savings achieved by the feedback-free OLRA scheme and reveal the tradeoff in terms of transmission reliability and latency when compared to the error-prone feedback-aware CLRA scheme. In essence, the results quantify the energy saving of the feedback-free OLRA scheme at the expense of transmission reliability reduction and latency increment compared with the feedback-aware CLRA scheme. 
%%%%%%%%%%%%%%%%%%%%%%%%%%%%%%%%%%%%%%%%%%%%%%%%%%%%
 % \vspace{-0.4cm}
\begin{appendices}
\section{Proof of Lemma 4}\label{app:lemma4}
The first and second moments of the FSD probability $\rm p_n$, i.e., $M_1$ and $M_2$ can be obtained by averaging $\rm p_n$ in (\ref{meta_def}) over the devices types, activities, fading gains, and the spatial locations of the devices. Hence, $M_1$ is given by
\begin{equation}\small
\begin{aligned}
	M_1 & = \mathbb{E}_{ h_o,\xi_i,g_i, \Psi} \left \{\frac{p_o h_o R_o^{-\eta}}{\sum_{v=1}^{V}\sum_{\rm x_i \in \Psi_v \setminus \rm x_0} \mathbbm{1}_{\xi_i} p_v g_i R_i ^{-\eta}} \geq \theta_n \right\} \nonumber \\
	% &= \mathbb{E}_{ h_o,\xi_i,g_i, \Psi} \left \{\mathbb{d}\left \{h_o \geq  \theta_n {\sum_{v=1}^{V}\sum_{\rm x_i \in \Psi_v \setminus \rm x_0} \mathbbm{1}_{\xi_i} \frac{p_v g_i R_0 ^{\eta}}{p_o R_i ^{\eta}}} \right\}\right\} \nonumber \\
	& \stackrel{(a)} = \mathbb{E}_{\xi_i, g_i, \Psi} \left\{\exp \left[-\theta_n {\sum_{v=1}^{V}\sum_{\rm x_i \in \Psi_v \setminus \rm x_0} \mathbbm{1}_{\xi_i} \frac{p_v g_i R_0 ^{\eta}}{p_o R_i ^{\eta}}} \right] \right\} \nonumber \\
	% & \stackrel{(b)} = \mathbb{E}_{g_i, \Psi} \left\{\prod_{v=1}^{V}\prod_{\rm x_i \in \Psi_v \setminus \rm x_0} \alpha_v \mathcal{L}_{g_i} \left(\frac{\theta_n p_v R_o^{\eta}}{p_o R_i^\eta}\right) +(1-\alpha_v) \right\} \nonumber \\
	& \stackrel{(b)} = \mathbb{E}_{\Psi} \left\{\prod_{v=1}^{V}\prod_{\rm x_i \in \Psi_v \setminus \rm x_0} \frac{\alpha_v}{1+ \frac{\theta_n p_v R_o^{\eta}}{p_o R_i^\eta} }+ (1-\alpha_v) \right\}\nonumber  \\
	& \stackrel{(c)} =  \exp \!\left[- 2 \pi \sum_{v=1}^{V} \lambda_v \!\! \int_{0}^{\infty} \!\!\! \left(1-\!\!\left(1-\frac{\alpha_v \frac{\theta_n p_v R_o^{\eta}}{p_o x^\eta}}{1+\frac{\theta_n p_v R_o^{\eta}}{p_o x^\eta}} \right) \right) \! \! x  \text{d} x \right]  \nonumber  \\
	% & =  \exp \left[- 2 \pi \sum_{v=1}^{V} \lambda_v \alpha_v \int_{x=0}^{\infty} \left(\frac{\frac{\theta_n p_v R_o^{\eta}}{p_o x^\eta}}{1+\frac{\theta_n p_v R_o^{\eta}}{p_o x^\eta}} \right) x  \text{d} x \right]  \nonumber  \\
	& \stackrel{(d)}  =  \exp \left[\frac{- 2 \pi R_o ^2 \theta_n^{2/\eta }} {\eta} \sum_{v=1}^{V} \left(\frac{p_v}{p_o}\right)^{\! \! \frac{2}{\eta}}\lambda_v  \alpha_v  \! \!\int_{0}^{\infty} \! \frac{y^{-\frac{2}{\eta}}}{1+y} \text{d} y \right ],
 % \nonumber  \\
	% & \stackrel{(e)}=  \exp  \left[\frac{- 2 \pi R_o ^2 \theta_n^{2/\eta }} {\eta} \sum_{v=1}^{V} \left(\frac{p_v}{p_o}\right)^{\! \! \frac{2}{\eta}}\lambda_v  \alpha_v   \mathcal{B} \left(-\frac{2}{\eta}+1,\frac{2}{\eta} \right) \right] \nonumber  \\
	% & \stackrel{(f)}= \exp \left[\frac{- 2 \pi^2  R_o ^2 }{\eta \sin \left(\frac{2 \pi}{\eta} \right)} \theta_n^{2/\eta } \sum_{v=1}^{V} \left(\frac{p_v}{p_o}\right)^{\! \! \frac{2}{\eta}} \lambda_v \alpha_v \right],
	%\label{1st_moment}
\end{aligned}
\end{equation}
where (a) follows from the exponentially-distributed power gain of the intended channel (i.e., $h_o \sim \exp (1) $); (b) follows from the Bernoulli distribution of the IoT interfering device activity indicator $\mathbbm{1}_{\xi_i}$ and $g_i \sim \exp (1) $; (c) follows from the probability generating functional (PGFL) of the PPP~\cite{haenggi2012stochastic}; (d) is obtained by changing variables $y=\frac{\theta_n p_v R_o^{\eta}}{p_o x^\eta}$. By manipulating the integral in (d), $M_1$ in~\eqref{1st_moment} is obtained.
% ; (e) is due to $\int_{0}^{\infty} \frac{x^{\mu-1}}{(1+\theta x)^v} \text{d}x = \theta^{-\mu} \, \mathcal{B} (\mu, v-\mu), \quad |\arg \theta| <\pi , \, \text{Re}[v]>\text{Re}[\mu]>0$; and (f) is obtained from $ \theta(x,1-x)=\frac{\pi}{\sin \left( \pi x \right)}$.\\

Similarly, the second moment of the FSD probability $M_2$ can be obtained as follows
\begin{equation}
 \begin{aligned}
	M_2 & = \mathbb{E}_{\Psi} \left[ \ \prod_{v=1}^{V}\prod_{\rm x_i \in \Psi_v \setminus \rm x_0} \left(\frac{\alpha_v}{1+ \frac{\theta_n p_v R_o^{\eta}}{p_o R_i^\eta} }+ (1-\alpha_v) \right)^2 \right] \nonumber\\
	% & = \mathbb{E}_{\Psi} \left[ \ \prod_{v=1}^{V}\prod_{\rm x_i \in \Psi_v \setminus \rm x_0} \left(1- \frac{\alpha_v\frac{\theta_n p_v R_o^{\eta}}{p_o R_i^\eta} }{1+ \frac{\theta_n p_v R_o^{\eta}}{p_o R_i^\eta} }\right)^2 \right]\nonumber\\
	% &=  \exp \left[- 2 \pi \sum_{v=1}^{V} \lambda_v \int_{x=0}^{\infty} \left[ 1- \left(1- \frac{\alpha_v\frac{\theta_n p_v R_o^{\eta}}{p_o R_i^\eta} }{1+ \frac{\theta_n p_v R_o^{\eta}}{p_o R_i^\eta} }\right)^2 \right] x  \text{d} x \right] \nonumber \\
	&= \exp \Bigg[- 2 \pi \sum_{v=1}^{V} \lambda_v \Bigg\{ 2\alpha_v \int_{x=0}^{\infty} \frac{\left(\frac{\theta_n p_v R_o^{\eta}}{p_o x ^\eta}\right)}{\left( 1+ \frac{\theta_n p_v R_o^{\eta}}{p_o x ^\eta} \right)} x \text{d}x \nonumber \\
	&\qquad \qquad \qquad \qquad -\alpha_v^2  \int_{x=0}^{\infty}  \frac{\left(\frac{\theta_n p_v R_o^{\eta}}{p_o x ^\eta}\right)^2}{\left( 1+ \frac{\theta_n p_v R_o^{\eta}}{p_o x^\eta} \right) ^2} x \text{d}x \Bigg\} \Bigg],
	%\label{2nd_moment}
\end{aligned}   
\end{equation}
By manipulating the integrals in a similar way to (\ref{1st_moment}), the expression of $M_2$ is obtained as provided in Lemma~\ref{Lemma: moments of meta distribution}.   

% {\footnotesize 
% \begin{align}
% 	& \int_{x=0}^{\infty}  \frac{\left(\frac{\theta_n p_v R_o^{\eta}}{p_o x ^\eta}\right)^2}{\left( 1+ \frac{\theta_n p_v R_o^{\eta}}{p_o x ^\eta} \right) ^2} x \text{d}x 
% 	= \left(\frac{\theta_n p_v }{p_o}\right)^{2/\eta} \frac{R_0^2}{\eta} \int_{y=0}^{\infty} \frac{y^{1-2/\eta}}{(1+y)^2} \text{d} y \nonumber \\
% 	&= \left(\frac{\theta_n p_v }{p_o}\right)^{\frac{2}{\eta}} \frac{R_0^2}{\eta} \, \mathcal{B}(2-2/\eta, 2/\eta) \stackrel{(a)} = \left(\frac{\theta_n p_v }{p_o}\right)^{\frac{2}{\eta}} \frac{R_0^2 (\eta-2) }{\eta^2} \frac{\pi}{\sin \left(\frac{2\pi}{\eta} \right)} 
% 	\label{I2}
% \end{align}}

% where (a) comes from $\mathcal{B} (x+1,y)=\mathcal{B} (x,y) \frac{x}{x+y}$. By substituting (\ref{I2}) into (\ref{2nd_moment}), $M_2$ expression in Lemma 1 is obtained.
\end{appendices}
%%%%%%%%%%%%%%%%%%%%%%%%%%%%%%%%%%%
\bibliographystyle{IEEEtran}
\bibliography{IEEEabrv,Ref}
\end{document}

%% file: defs_tikzpgf.tex
\usepackage{tikz}
\usepackage{pgfplotstable}
\usepackage{rotate}

\definecolor{rubblue}{cmyk}{1,0.5,0,0.6}
\definecolor{rubgreen}{cmyk}{0.5,0,1,0}
\definecolor{rubgray}{cmyk}{0.03,0.03,0.03,0.1}

\usepgfplotslibrary{units}

\usetikzlibrary{%
patterns,%
calc,%
fit,%
arrows,%
plotmarks,%
shadows,%
chains,%
shapes%
}

\tikzset{>=latex'} % requires arrows-library
\tikzstyle{every picture}+=[remember picture] % overlays
\pgfdeclarelayer{background}
\pgfdeclarelayer{foreground}
\pgfsetlayers{background,main,foreground}

\tikzstyle{blueblock}=[draw=rubblue, rectangle, thick, drop shadow, minimum width=20mm, minimum height=8mm,fill=rubblue!20, text width=20mm, text centered]
\tikzstyle{bluebox}=[draw=rubblue, rectangle, thick, drop shadow, minimum width=8mm, minimum height=8mm,fill=rubblue!20, text width=8mm, text centered]%, rounded corners=3pt]
\tikzstyle{greenblock}=[draw=rubgreen, rectangle, thick, drop shadow, minimum width=20mm, minimum height=8mm,fill=rubgreen!20, text width=20mm, text centered]
\tikzstyle{dot} = [draw, circle, minimum size=0.2pt,scale=0.3,fill=black,black]
\tikzstyle{smalldot} = [draw, circle, minimum size=0.1pt,scale=0.2,fill=black,black]
\tikzstyle{reddot}  =[draw,circle,minimum size=0.2pt,scale=0.8,fill=red,thin]
\tikzstyle{greendot}  =[draw,circle,minimum size=0.2pt,scale=0.8,fill=Green,thin]
\tikzstyle{bluedot}  =[draw,circle,minimum size=0.2pt,scale=0.8,fill=blue,thin]
\tikzstyle{whitedot}=[draw,circle,minimum size=0.2pt,scale=0.8,fill=white,thin]
\tikzstyle{blackdot} = [draw, circle, minimum size=0.2pt,scale=0.7,fill=black,black]
\tikzstyle{sum} = [drop shadow, draw=rubblue, thick, fill=rubblue!20, circle]
\tikzstyle{relay} = [blueblock, minimum width=5mm, minimum height=20mm, text width=5mm, rounded corners=2pt]
\tikzstyle{relay2} = [blueblock, minimum width=5mm, minimum height=15mm, text width=5mm, rounded corners=2pt]
\tikzstyle{relay3} = [blueblock, minimum width=5mm, minimum height=25mm, text width=5mm, rounded corners=2pt]
\tikzstyle{relay4} = [blueblock, minimum width=5mm, minimum height=10mm, text width=5mm, rounded corners=2pt]
\tikzstyle{relay5} = [blueblock, minimum width=5mm, minimum height=50mm, text width=5mm, rounded corners=2pt]
\tikzstyle{relay6} = [blueblock, minimum width=5mm, minimum height=5mm, text width=5mm, rounded corners=2pt]
\tikzstyle{circgreen} = [draw, circle, inner sep=2pt, fill=rubgreen, drop shadow, thick]
\tikzstyle{circwhite} = [draw, circle, inner sep=2pt, fill=white, drop shadow, thick]
\tikzstyle{circdashed} = [draw, dashed, circle, inner sep=2pt, fill=rubgray, drop shadow, thick]
\tikzstyle{vertbox} = [rectangle, draw=rubblue, thick, rotate=90, text centered, minimum width=16.5mm, minimum height=8mm, text width=16.5mm, inner sep=0pt, fill=rubblue!20, drop shadow]
\tikzstyle{vertboxb} = [rectangle, draw=rubblue, thick, rotate=90, text centered, minimum width=16.5mm, minimum height=8mm, text width=16.5mm, fill=rubblue!20, drop shadow]
\tikzstyle{vertboxshort} = [rectangle, draw=rubblue, thick, rotate=90, text centered, minimum width=10mm, minimum height=8mm, text width=10mm, inner sep=0pt, fill=rubblue!20, drop shadow]
\tikzstyle{smalldotgreen} = [draw=rubgreen, circle, minimum size=0.2pt,scale=0.8,fill=rubgreen!20]
\tikzstyle{antenna} = [regular polygon, regular polygon sides=3, draw, shape border rotate=180, minimum size=0.2pt, scale=0.3]

\tikzstyle{poly} = [regular polygon, regular polygon sides=6, shape aspect=0.5, minimum width=1.5cm, minimum height=0.35cm, draw, dashed]

\definecolor{cff9e00}{RGB}{255,158,0}
\definecolor{c4fff00}{RGB}{79,255,0}
\definecolor{cff0012}{RGB}{255,0,18}
\definecolor{c00c5ff}{RGB}{0,197,255}
\definecolor{c046f00}{RGB}{4,111,0}
\definecolor{c004b9d}{RGB}{0,75,157}
\newlength{\mylen}

%% file: MC_CLRA.tikz
			\begin{tikzpicture}[,scale=0.375, align=center,
				background rectangle/.style={draw=black!10,fill=black!15,rounded corners=1ex,  fill opacity=0.4},
				background rectangle1/.style={draw=black!10,fill=yellow!30,rounded corners=1ex,  fill opacity=0.4},
				idle_node/.style={circle, draw=black!60, fill=magenta!25, thin, minimum size=8mm},
				a_node/.style={circle, draw=black!60, fill=green!15, thin, minimum size=8mm},
				b_node/.style={circle, draw=black!60, fill=blue!15, thin, minimum size=8mm},
				c_node/.style={circle, draw=black!60, fill=red!15, thin, minimum size=8mm},
				absorb_node/.style={circle, draw=black!60, fill=cyan!15, thin, minimum size=10mm}
				]
				
				%Nodes
				\node[idle_node]     (idle) at(-4,0)    {$\mathbf{0}$};
				
				\node[a_node]        (a1)   at(0,0)    {\textbf{a}};
				\node[a_node]        (a2)   at(4,0)    {\textbf{a}};
				\node[a_node]        (a3)   at(8,0)    {\textbf{a}};
				\node[a_node]        (a4)   at(12,0)   {\textbf{a}};
				\node[a_node]        (a5)   at(16,0)   {\textbf{a}};
				\node[a_node]        (a6)   at(20,0)   {\textbf{a}};

				\node[b_node]        (b1)   at(4,-3) {\textbf{b}};
				\node[b_node]        (b2)   at(8,-3) {\textbf{b}};
				\node[b_node]        (b3)   at(12,-3) {\textbf{b}};
				\node[b_node]        (b4)   at(16,-3) {\textbf{b}};
				\node[b_node]        (b5)   at(20,-3) {\textbf{b}};
				\node[b_node]        (b6)   at(24,-3) {\textbf{b}};
				
				\node[c_node]        (c1)   at(8,-6) {\textbf{c}};
				\node[c_node]        (c2)   at(12,-6) {\textbf{c}};
				\node[c_node]        (c3)   at(16,-6) {\textbf{c}};
				\node[c_node]        (c4)   at(20,-6) {\textbf{c}};
				\node[c_node]        (c5)   at(24,-6) {\textbf{c}};
				\node[c_node]        (c6)   at(28,-6) {\textbf{c}};
				
				\node[absorb_node]      (s)    at(34,-6) {\textbf{S}};
				\node[absorb_node]      (to)   at(34,0) {\textbf{t-out}};
				
				%Header
				\node at(0,3) {$\mathbf{t=1}$};
				\node at (4,3) {$\mathbf{t=2}$}; 
				\node at (8,3) {$\mathbf{t=3}$}; 
				\node at (12,3) {$\mathbf{t=4}$}; 
				\node at (16,3) {$\mathbf{t=5}$}; 
				\node at (20,3) {$\mathbf{t=6}$}; 
				\node at (24,3) {$\mathbf{t=7}$}; 
				\node at (28,3) {$\mathbf{t=8}$};
				\node at (34,1) [label={\textbf{ \small Absorbing} \\ \textbf{\small states}}] (one){};
				\node at (-4,1) [label={\textbf{ \small Idle} \\ \textbf{\small state}}] (one){};
				
				%\node at (8,4.5) [text = red] {\tiny $\mathbf{t=n}$};
				%			\node at (20,4.5) [text = red] {\tiny $\mathbf{t=T-n+1}$};
				
				% Rectangles
				\draw [background rectangle] (-5.5,4) rectangle (-2.5,-2);
				\draw [background rectangle] (-1.5,4) rectangle (1.5,-2);
				\draw [background rectangle] (2.5,4) rectangle (5.5,-5);
				\draw [background rectangle1] (6.5,4) rectangle (9.5,-8);
				\draw [background rectangle] (10.5,4) rectangle (13.5,-8);
				\draw [background rectangle] (14.5,4) rectangle (17.5,-8);
				\draw [background rectangle1] (18.5,4) rectangle (21.5,-8);
				\draw [background rectangle] (22.5,4) rectangle (25.5,-8);
				\draw [background rectangle] (26.5,4) rectangle (29.5,-8);
				\draw [background rectangle] (31.5,4) rectangle (37.5,-8);
				
				%Lines
				\draw[dashed,line width=0.4mm, ->] (a1.east) -- (a2.west);
				\draw[dashed,line width=0.4mm, ->] (a2.east) -- (a3.west);
				\draw[dashed,line width=0.4mm, ->] (a3.east) -- (a4.west);
				\draw[dashed,line width=0.4mm, ->] (a4.east) -- (a5.west);
				\draw[dashed,line width=0.4mm, ->] (a5.east) -- (a6.west);
				
				\draw[dashed,line width=0.4mm, ->] (b1.east) --  (b2.west);
				\draw[dashed,line width=0.4mm, ->] (b2.east) -- (b3.west);
				\draw[dashed,line width=0.4mm, ->] (b3.east) -- (b4.west);
				\draw[dashed,line width=0.4mm, ->] (b4.east) -- (b5.west);
				\draw[dashed,line width=0.4mm, ->] (b5.east) -- (b6.west);
				\draw[dashed,line width=0.4mm, ->] (c1.east) -- (c2.west);
				\draw[dashed,line width=0.4mm, ->] (c2.east) -- (c3.west);
				\draw[dashed,line width=0.4mm, ->] (c3.east) -- (c4.west);
				\draw[dashed,line width=0.4mm, ->] (c4.east) -- (c5.west);
				\draw[dashed,line width=0.4mm, ->] (c5.east) --  (c6.west);
				
				\draw[line width=0.4mm, ->] (a1.320) -- (b1.140);
				\draw[line width=0.4mm, ->] (a2.320) --  (b2.140);
				\draw[line width=0.4mm, ->] (a3.320) --  (b3.140);
				\draw[line width=0.4mm, ->] (a4.320) --  (b4.140);
				\draw[line width=0.4mm, ->] (a5.320) --  (b5.140);
				\draw[line width=0.4mm, ->] (a6.320) --  (b6.140);
				
				\draw[line width=0.4mm, ->] (b1.320) --  (c1.140);
				\draw[line width=0.4mm, ->] (b2.320) --  (c2.140);
				\draw[line width=0.4mm, ->] (b3.320) --  (c3.140);
				\draw[line width=0.4mm, ->] (b4.320) --  (c4.140);
				\draw[line width=0.4mm, ->] (b5.320) --  (c5.140);
				\draw[line width=0.4mm, ->] (b6.320) --  (c6.140);
				
				\draw[line width=0.4mm, red, ->] (idle.east) -- (a1.west);
				
				% absorbing:
				\def\myshift#1{\raisebox{-2.5ex}}
				\draw [->,,line width=0.4mm,postaction={decorate}] (c1.280) to [bend right=30]  (s.south);
				\draw [->,,line width=0.4mm,postaction={decorate}] (c2.280) to [bend right=30]  (s.240);
				\draw [->,,line width=0.4mm,postaction={decorate}] (c3.280) to [bend right=30]  (s.225);
				\draw [->,,line width=0.4mm,postaction={decorate}] (c4.280) to [bend right=30]  (s.200);
				\draw [->,,line width=0.4mm,postaction={decorate}] (c5.280) to [bend right=30]  (s.180);
				\draw[line width=0.4mm, ->] (c6.east) -- (s.west);
				
				\draw [->,,dashed,line width=0.4mm,postaction={decorate}] (b6.20) to [bend right=10]  (to.200);
				\draw[dashed,line width=0.4mm, ->] (a6.east) --  (to.west);
				\draw [->,,dashed,line width=0.4mm,postaction={decorate}] (c6.20) to [bend right=10]  (to.250);
				
				\draw[line width=0.4mm, red, ->] (-5,-6.5) -- (-3,-6.5) node at (0,-6.5) {$1$};
				
				\draw[line width=0.4mm, ->] (-5,-8) -- (-3,-8) node at (1,-8) {$\rho=\rm p_{n,m} \, \rm p_{ack}$};
				\draw[dashed, line width=0.4mm, ->] (-5,-9.5) -- (-3,-9.5) node at (1,-9.5) {$\bar{\rho}=1-\rho$};
				\draw (s) edge  [loop right,line width=0.4mm, red]( );
				\draw (to) edge  [loop right,line width=0.4mm, red]( );
				\draw [line width=0.4mm](-5.5,-10.5) rectangle (4.5,-5.5);
			\end{tikzpicture}

%% file: MC_OLRA.tikz
\begin{tikzpicture}[thin, scale=0.75, align=center,
				background rectangle/.style={draw=black!10,fill=black!15,rounded corners=1ex,  fill opacity=0.4},
				idle_node/.style={circle, draw=black!60, fill=magenta!25, thin, minimum size=10mm},
				a_node/.style={circle, draw=black!60, fill=green!15, thin, minimum size=10mm},
				b_node/.style={circle, draw=black!60, fill=blue!15, thin, minimum size=10mm},
				c_node/.style={circle, draw=black!60, fill=red!15, thin, minimum size=10mm},
				absorb_node/.style={circle, draw=black!60, fill=cyan!15, thin, minimum size=12mm},
				stay_node/.style={circle, draw=black!60, fill=yellow!15, thin, minimum size=10mm}
				]
				
				%Nodes
				\node[idle_node]     (idle) at(-4,0)    {0};
				
				\node[a_node]        (a1)   at(-2,0)    {$\mathbf{a_1}$};
				\node[a_node]        (a2)   at(0,0)    {$\mathbf{a_2}$};
				\node[a_node]        (a3)   at(2,0)    {$\mathbf{a_3}$};
				
				\node[a_node]        (a4)   at(4,0)    {$\mathbf{a_4}$};
				
				\node[b_node]        (b1)   at(6,0) {$\mathbf{b_1}$};
				\node[b_node]        (b2)   at(8,0) {$\mathbf{b_2}$};
				\node[b_node]        (b3)   at(10,0) {$\mathbf{b_3}$};
				
				\node[c_node]        (c1)   at(12,0) {$\mathbf{c_1}$};
				\node[c_node]        (c2)   at(14,0) {$\mathbf{c_2}$};
				\node[c_node]        (c3)   at(16,0) {$\mathbf{c_3}$};
				\node[c_node]        (c4)   at(18,0) {$\mathbf{c_4}$};    
				
				\node[absorb_node]   (s)    at(21,1.5) {\textbf{S}};
				\node[absorb_node]   (to)   at(21,-2) {\textbf{t-out}};
				
				\node[stay_node]     (LOG1) at (0,2) {\textbf{LS}};  
				\node[stay_node]     (LOG2) at (2,2) {\textbf{LS}};
				
				\node[stay_node]     (LOG3) at (4,2) {\textbf{LS}};
				
				\node[stay_node]     (LOG4) at (8,2) {\textbf{LS}};
				\node[stay_node]     (LOG5) at (10,2) {\textbf{LS}};

				%Header
				\node at(-2,3.5) {   $\mathbf{t=1}$};
				\node at (0,3.5) { $\mathbf{t=2}$}; 
				\node at (2,3.5) { $\mathbf{t=3}$}; 
				\node at (4,3.5) { $\mathbf{t=4}$}; 
				\node at (6,3.5) { $\mathbf{t=5}$}; 
				\node at (8,3.5) { $\mathbf{t=6}$}; 
				\node at (10,3.5) {$\mathbf{t=7}$}; 
				\node at (12,3.5) { $\mathbf{t=8}$};
				\node at (14,3.5) { $\mathbf{t=9}$};
				\node at (16,3.5) { $\mathbf{t=10}$};
    		\node at (18,3.5) { $\mathbf{t=11}$};
				
				\node at (21,2.5) [label={\textbf{  Absorbing} \\ \textbf{  states}}] (one){};
				\node at (-4.25,0.6) [label={\textbf{    Idle} \\ \textbf{  state}}] (one){};
				
				\node at (1,-3.8) [label={\textbf{   Fragment $a$ is sent $\kappa+1$ times }}] (one){};
				\node at (8,-3.8) [label={  \textbf{  Fragment $b$ is sent $\kappa$ times }}] (one){};
				\node at (15,-3.8) [label={\textbf{  Fragment $c$ is sent $\kappa+1$ times }}] (one){};
				
				% Rectangles
				\draw [background rectangle] (-5,2) rectangle (-3.2,-1);
				\draw [background rectangle] (-2.9,4) rectangle (4.8,-4);
				\draw [background rectangle] (5.1,4) rectangle (10.85,-4);
				\draw [background rectangle] (11.15,4) rectangle (18.85,-4);
				\draw [background rectangle] (19.5,4) rectangle (23,-4);

				% Lines
				\draw[line width=0.5mm, red, ->] (idle.east) -- (a1.west);
				
				\draw[dashed,line width=0.5mm, ->] (a1.east) --  (a2.west);
				\draw[dashed,line width=0.5mm, ->] (a2.east) -- (a3.west);
				\draw[dashed,line width=0.5mm, ->] (a3.east) -- (a4.west);
				
				\draw[dashed,line width=0.5mm, ->] (b1.east) --  (b2.west);
				\draw[dashed,line width=0.5mm, ->] (b2.east) --  (b3.west);
				
				\draw[dashed,line width=0.5mm, ->] (c1.east) --  (c2.west);
				\draw[dashed,line width=0.5mm, ->] (c2.east) --  (c3.west);	
				\draw[dashed,line width=0.5mm, ->] (c3.east) --  (c4.west);	
    
				\draw[line width=0.5mm, ->] (a1.40) -- (LOG1.200);
				\draw[line width=0.5mm, ->] (a2.30) --  (LOG2.220);			
				\draw[line width=0.5mm, ->] (a3.30) --  (LOG3.220);	
				
				\draw[line width=0.5mm, ->] (b1.40) -- (LOG4.200);
				\draw[line width=0.5mm, ->] (b2.30) --  (LOG5.220);

				\draw[line width=0.5mm, red, ->] (LOG1.east) -- (LOG2.west);
				\draw[line width=0.5mm, red, ->] (LOG2.east) --(LOG3.west);
				\draw[line width=0.5mm, red, ->] (LOG4.east) --(LOG5.west);
				
				\draw[line width=0.5mm, red, ->] (LOG3.east) -- (b1.100);
				\draw[line width=0.5mm, red, ->] (LOG5.east) -- (c1.100);
				
				\draw[line width=0.5mm, ->] (a4.east) --(b1.west);
				\draw[line width=0.5mm, ->] (b3.east) -- (c1.west);

				% absorbing:
				\def\myshift#1{\raisebox{-2.5ex}}
				\draw [->,,dashed,line width=0.5mm] (a4.300) to [bend right=10]  (to.200);
				
				\draw [->,,dashed,line width=0.5mm] (b3.300) to [bend right=10]  (to.160);            
				
				% \draw [->,,dashed,line width=0.5mm] (c3.320) to [bend right=10]  (to.130);
    
                 \draw [->,,dashed,line width=0.5mm] (c4.320) to [bend right=10]  (to.100);
				
				\draw [->,,line width=0.5mm] (c1.80) to [bend left=20]  (s.150);
				
				\draw [->,,line width=0.5mm] (c2.80) to [bend left=20]  (s.170);            
				
				\draw [->,,line width=0.5mm] (c3.80) to [bend left=20]  (s.190);
    		
                \draw [->,,line width=0.5mm] (c4.80) to [bend left=20]  (s.210);
				
				\draw (s) edge  [loop right, red,line width=0.5mm]( );
				\draw (to) edge  [loop right, red,line width=0.5mm]( );
				
				\draw[line width=0.5mm, red, ->] (-5.75,-1.75) -- (-4.75,-1.75) node at (-4,-1.75) {$ {1}$};
				
				\draw[line width=0.5mm, ->] (-5.75,-2.75) -- (-4.75,-2.75) node at (-4,-2.75) {$  {\rm p_{n,m}}$};
				
				\draw[dashed, line width=0.5mm, ->] (-5.75,-3.75) -- (-4.75,-3.75) node at (-4,-3.75) {$  {\rm{\bar p}_{n,m}}$};
				
				\draw [line width=0.5mm](-6,-4.25) rectangle (-3.25,-1.25);
			\end{tikzpicture}